%% file: ms.tex
\documentclass[twocolumn]{aastex631}

\usepackage{showyourwork}
\usepackage{bm}
\usepackage{amsmath}
\usepackage{appendix}
\input{macros.tex}
\newcommand{\CIPlusMinus}[1]{{#1[median]^{+#1[error plus]}_{-#1[error minus]}}}

\newcommand{\msun}{\ensuremath{{\rm M}_\odot}}
\newcommand{\result}[1]{#1}

\newcommand{\NewChange}[1]{#1}
\newcommand{\LinkExplainer}{The icons link to the code used to generate this figure and to \href{https://zenodo.org}{zenodo} entries of any public data used.}

\graphicspath{{./}{./figures/}}

\begin{document}

\title{Cover Your Basis: Comprehensive Data-Driven Characterization of the Binary Black Hole Population}

\author{Bruce Edelman}
\email{bedelman@uoregon.edu}
\affiliation{Institute  for  Fundamental  Science, Department of Physics, University of Oregon, Eugene, OR 97403, USA}
\author{Ben Farr}
\affiliation{Institute  for  Fundamental  Science, Department of Physics, University of Oregon, Eugene, OR 97403, USA}
\author{Zoheyr Doctor}
\affiliation{Center for Interdisciplinary Exploration and Research in Astrophysics (CIERA), Department of Physics and Astronomy, Northwestern
University, Evanston, IL 60201, USA}

\begin{abstract}                 
We introduce the first complete non-parametric model for the astrophysical distribution of the binary black hole (BBH) population. Constructed from basis splines, we use these models to conduct the most comprehensive data-driven investigation of the BBH population to date, simultaneously fitting non-parametric models for the BBH mass ratio, spin magnitude and misalignment, and redshift distributions. With GWTC-3, we report the same features previously recovered with similarly flexible models of the mass distribution, most notably the peaks in merger rates at primary masses of ${\sim}10\msun$ and ${\sim}35\msun$. Our model reports a suppressed merger rate at low primary masses and a mass ratio distribution consistent with a power law. We infer a distribution for primary spin misalignments that peaks away from alignment, supporting conclusions of recent work. We find broad agreement with the previous inferences of the spin magnitude distribution: the majority of BBH spins are small ($a<0.5$), the distribution peaks at $a\sim0.2$, and there is mild support for a non-spinning subpopulation, which may be resolved with larger catalogs. With a modulated power law describing the BBH merger rate's evolution in redshift, we see hints of the rate evolution either flattening or decreasing at $z\sim0.2-0.5$, but the full distribution remains entirely consistent with a monotonically increasing power law. We conclude with a discussion of the astrophysical context of our new findings and how non-parametric methods in gravitational-wave population inference are uniquely poised to complement to the parametric approach as we enter the data-rich era of gravitational-wave astronomy.
\end{abstract}

\input{intro.tex}

\input{methods.tex}

\input{results.tex}

\input{conclusion.tex}

\section{Acknowledgements}\label{sec:acknowledments}
We thank Tom Callister, Will Farr, Maya Fishbach, Salvatore Vitale, and Jaxen Godfrey for useful discussions during the 
preparation of this manuscript and/or helpful comments on early drafts. Z.D. also acknowledges support from 
the CIERA Board of Visitors Research Professorship.  This research has made use of data, software and/or web tools obtained 
from the Gravitational Wave Open Science Center (\url{https://www.gw-openscience.org/}), a service of LIGO Laboratory, the 
LIGO Scientific Collaboration and the Virgo Collaboration. The authors are grateful for computational resources 
provided by the LIGO Laboratory and supported by National Science Foundation Grants PHY-0757058 and PHY-0823459.  
This work was supported in part by the National Science Foundation under Grant PHY-2146528 and benefited from access 
to the University of Oregon high performance computer, Talapas. This material is based upon work supported in part 
by the National Science Foundation under Grant PHY-1807046 and work supported by NSF's LIGO Laboratory which is a major 
facility fully funded by the National Science Foundation.

\software{
\textsc{ShowYourWork}~\citep{Luger2021},
\textsc{Matplotlib}~\citep{Hunter:2007},
\textsc{NumPy}~\citep{harris2020array},
\textsc{SciPy}~\citep{2020SciPy-NMeth},
\textsc{AstroPy}~\citep{2018AJ....156..123A},
\textsc{Jax}~\citep{jax},
\textsc{NumPyro}~\citep{pyro,numpyro},
}

\bibliography{bib}{}
\bibliographystyle{aasjournal}

\input{appendicies.tex}

\end{document}

%% file: intro.tex
\section{Introduction} \label{sec:intro}

Observations of gravitational waves (GWs) from compact binary mergers are becoming a regular occurrence, 
producing a catalog of events that recently surpassed 90 such detections \citep{GWTC1,gwtc2,GWTC3}. As the catalog continues to grow, so does our understanding of the underlying astrophysical population of compact binaries \citep{o1o2_pop,o3a_pop,o3b_astro_dist}. 
Following numerous improvements to the detectors since the last observing run, the anticipated sensitivities for the upcoming fourth observing run of the LIGO-Virgo-KAGRA (LVK) collaboration suggest detection rates as high as once per \emph{day} \citep{aLIGO,aVIRGO,LVK_prospects,KAGRAProg}. With the formation history of these dense objects encoded in the details of their distribution \citep{1503.04307,Rodriguez_2016,Farr2017Nature,Zevin_2017,Farr_BinnedSpin}, the likely doubling in size of the catalog with the next observing run could provide another leap in our understanding of compact binary astrophysics. 
Beyond formation physics, population-level inference of the compact binary catalog has also been shown to provide novel measurements of cosmological parameters \citep{Farr_2019HUB,gwtc3_cosmo,JoseSpectralSirens}, constrain modified gravitational wave propagation \citep{OkounkovaBirefringence,ModGWProp,ModGWProp2}, 
constrain a running Planck mass \citep{Lagos_runningPlanckMass}, search for evidence of ultralight bosons through superradiance \citep{Ng_Boson2021,GWTC2_superradiance_Ng}, 
constrain stellar nuclear reaction rates \citep{Farmer_2019,Farmer_2020}, look for primordial black holes \citep{Ng_2021,KenNgPBH2022}, 
and to constrain physics of neutron stars \citep{Golomb_EOS,LandryRead_NS_Masses2021}. Through a better understanding of the mass, spin, and redshift distributions of 
compact binaries that will come with the increased catalog size, one can probe a wide range of different physical phenomena with even greater fidelity.

The binary black hole (BBH) mass distribution was first found to have structure beyond a smooth power law with simpler parametric models, exhibiting a possible high mass truncation and either a break or a peak at $m_1\sim35-40\,\msun$ \citep{Fishbach_2017,Talbot_2018,o1o2_pop,o3a_pop}. Starting with the moderately sized catalog, GWTC-2, more flexible models found signs of additional structure \citep{Tiwari_2021_b,Edelman_2022ApJ}. The evidence supporting these features, 
such as the peak at $m_1\sim10\,\msun$, has only grown after analyzing the latest catalog, GWTC-3, with the same models \citep{o3b_astro_dist,Tiwari_2022ApJ}. 
While this shows the usefulness of data-driven methods with the current relatively small catalog size, they will become more powerful with more observations. The canonical approach to constructing population models has been 
to use simple parametric descriptions (e.g., power laws, beta distributions) that aim to describe the data in the simplest way, employ astrophysically motivated priors where appropriate, then sequentially add 
complexities (e.g., Gaussian peaks) as the data demands. This simple approach was necessary when data was scarce, but as we move into the data-rich catalog era, this approach is already failing to scale.  More flexible and scalable methods, such as the non-parametric modeling techniques presented in this manuscript, will be necessary to continue to extract the full information contained in the compact binary catalog. In contrast to parametric models, flexible and non-parametric models are data-driven and contribute little bias to functional form. They additionally are particularly useful to search for unexpected features in the 
data, providing meaningful insight into features that parametric models may fail to capture.

While we eventually hope to uncover hints of binary formation mechanisms in the mass spectrum of BBHs, the distribution of spin properties have been of particular interest.  
The measurement of spin properties of individual binaries often have large uncertainties, but the theorized formation channels are expected to produce distinctly 
different spin distributions \citep{Rodriguez_2016,Gerosa2018, Farr_BinnedSpin, Farr2017Nature, Zevin_2017}. Isolated (or field) formation scenarios predict component spins that are preferentially aligned with the binary's orbital angular momentum, although some small 
misalignment can occur depending on the nature of the supernova kicks as each star collapses to a compact object \citep{Zevin_2022, BaveraBBHSpin, BaveraMassTransfer}. Alternatively, dynamical formation in 
dense environments where many-body interactions between compact objects can result in binary formation and hardening (shrinking of binary orbits) should produce binaries with components' spins distributed isotropically \citep{Rodriguez_2016, Rodriguez_2019}. BBH spins have also been of controversial interest recently, with different parametric approaches to modeling 
the spin distribution coming to different conclusions. Studies have disagreed on the possible existence of a significant zero-spinning subpopulation, as well as the presence of 
significant spin misalignment (i.e. $\cos{\theta_i} < 0.0$) \citep{o3b_astro_dist,RouletGWTC2Pop,BuildBetterSpinModels,GWTC3MonashSpin,Callister_NoEvidence}. 
Another study recently showed that inferences of spin misalignment (or tilts) are sensitive to modeling choices and may not peak at perfectly aligned spins, as is often assumed \citep{spinitasyoulike}. While enlightening,
these recent efforts to improve BBH spin models continue to build sequentially on previous parametric descriptions \citep{BuildBetterSpinModels,Callister_NoEvidence,spinitasyoulike}. 
To ensure we are extracting the full detail the catalog has to offer, we extend our previous non-parametric modeling techniques to include spin magnitudes and tilts, as well as the binary mass ratio and redshift. \NewChange{\citet{2210.12287} was released concurrently with this work (based on our previous work \citet{Edelman_2022ApJ}), and find similar conclusions on the spin distribution when applying similar flexible models constructed with cubic splines. The work presented in this manuscript however, does not need to analyze a suite of different model configurations and includes flexible non-parametric models for each of the mass, spin and redshift distributions rather than spin alone.}

Polynomial splines have been applied to success across different areas of gravitational-wave astronomy. They have been used to model the gravitational-wave data noise spectrum, 
detector calibration uncertainties, coherent gravitational waveform deviations, and modulations to a power law mass distribution \citep{Littenberg_2015,Edwards_2018,B_Farr_etal_2014,Edelman_2021,Edelman_2022ApJ}
In this paper we highlight how the use of basis-splines can provide a powerful non-parametric modeling approach to the astrophysical distributions of compact 
binaries. We illustrate how one can efficiently model both the mass and spin distributions of merging compact binaries in GWTC-3 with basis splines to infer compact binary population properties using 
hierarchical Bayesian inference. We discuss our results in the context of current literature on compact object populations and how this method complements the simpler lower 
dimensional parametric models in the short run, and will become necessary with future catalogs. Should they appear with more observations, this data-driven approach will provide checks of 
our understanding by uncovering more subtle -- potentially unexpected -- features. The rest of this manuscript is structured as follows: a description of the background of 
basis splines in section \ref{sec:basis_splines}, followed by a presentation of the results of our extensive, data-driven study of the mass and spin distributions of BBHs in GWTC-3 in section 
\ref{sec:results}. We then discuss these results and their astrophysical implications in section \ref{sec:astrodiscussion} and finish with our conclusions in section \ref{sec:conclusion}.

%% file: methods.tex
\section{Building the Model} \label{sec:methods}

We construct our data-driven model with the application of basis splines, or B-Splines \citep{deBoor78}. B-Splines of order $k$ are a set of order $k$ polynomials that 
span the space of possible spline functions interpolated from a given set of knot locations. For all B-Splines used in our model we use a third order basis which consists of individual cubic polynomials. The basis representation of the splines allows for the computationally
expensive interpolation to be done in a single preprocessing step -- amortizing the cost of model evaluation during inference. To mitigate the unwanted side effects of 
adding extra knots and to avoid running a model grid of differing numbers of knots (as in \citet{Edelman_2022ApJ}), we use the smoothing prior for Bayesian P-Splines \citep{eilers2021practical,BayesianPSplines,Jullion2007RobustSO}, 
allowing the data to pick the optimal scale needed to fit the present features. We discuss basis splines, the smoothing prior, and our specific prior choices on hyperparameters in Appendix \ref{sec:basis_splines}, \ref{sec:psplines} and \ref{sec:modelpriors}.

We parameterize each binaries' masses with the primary (more massive component) mass ($m_1$) 
and the mass ratio ($q=m_2/m_1$) with support from 0 to 1. Furthermore, we model 4 of the 6 total 
spin degrees of freedom of a binary merger: component spin magnitudes $a_1$ and $a_2$, and (cosine of) the tilt angles of each component, $\cos{\theta_1}$ and $\cos{\theta_2}$. The tilt angle is defined as the angle between each components' spin vector and the binary's orbital angular momentum vector. 
We assume the polar spin angles are uniformly distributed in the orbital plane. For the primary mass distribution, we model the log probability with a B-Spline interpolated over 
knots linearly spaced in $\log(m_1)$ from a minimum black hole mass, which we fix to $5\msun$, and a maximum mass 
that we set to $100\msun$. We then have the hyper-prior on primary mass with log probability density 
$\log(p(m_1 | \bm{c})) \propto B_{k=3}(\log(m_1) | \bm{c})$, where $B_{k=3}$ is the 
cubic B-Spline function with a vector of basis coefficients $\bm{c}$. We follow the same procedure 
for the models in mass ratio and spin distributions with knots spaced linearly across each domain 
so that we have $\log(p(\theta | \bm{c}_\theta)) \propto B_{k=3}(\theta | \bm{c}_\theta)$, 
where $\theta$ can be $q$, $a_1$, $a_2$, $\cos{\theta_1}$ or $\cos{\theta_2}$. 
For the spin magnitude and tilt distributions we construct two versions of the model: first, we model  
each component's distribution as independently and identically distribution (IID), 
where we have a single B-Spline model and parameters (coefficients) for each binary spin. 
Secondly, we model each component's distribution to be unique, fitting separate sets of coefficients for the B-Spline models of the primary and secondary spin distributions. 
Lastly, we fit a population model on the redshift or luminosity distance distribution of BBHs, assuming a $\Lambda\mathrm{CDM}$ cosmology defined by the parameters 
from the Planck 2015 results \citep{Planck2015}. This defines an analytical mapping between each event's inferred luminosity distance, and its redshift, which we now use interchangeably. 
We take a semi-parametric approach to model the evolution of the merger rate with redshift, following \citet{Edelman_2022ApJ}, that parameterizes modulations to an underlying model 
with splines (in our case basis splines). We use the \textsc{PowerlawRedshift} model as the underlying distribution to modulate, which has a single hyperparameter, $\lambda_z$, and 
probability density defined as: $p(z|\lambda_z)\propto \frac{dV_c}{dz}(1+z)^{\lambda_z-1}$ \citep{Fishbach_2018redshift}. For more detailed descriptions of each model and 
specific prior choices used for the hyperparameters see Appendix \ref{sec:modelpriors}. Now that we have our comprehensive data-driven population model built, we simultaneously fit the basis spline models on the BBH masses, spins and redshift. We use the usual hierarchical Bayesian inference framework (see appendix \ref{sec:hierarchical_inference} for a review; \citet{o1o2_pop,o3a_pop,o3b_astro_dist}), to perform the most 
extensive characterization of the population of BBHs to date using the most recent catalog of GW observations, GWTC-3 \citep{GWTC3}.

%% file: results.tex
\section{Binary Black Hole Population Inference with GWTC-3} \label{sec:results}

\begin{figure*}[ht!]
    \begin{centering}
        \includegraphics[width=\linewidth]{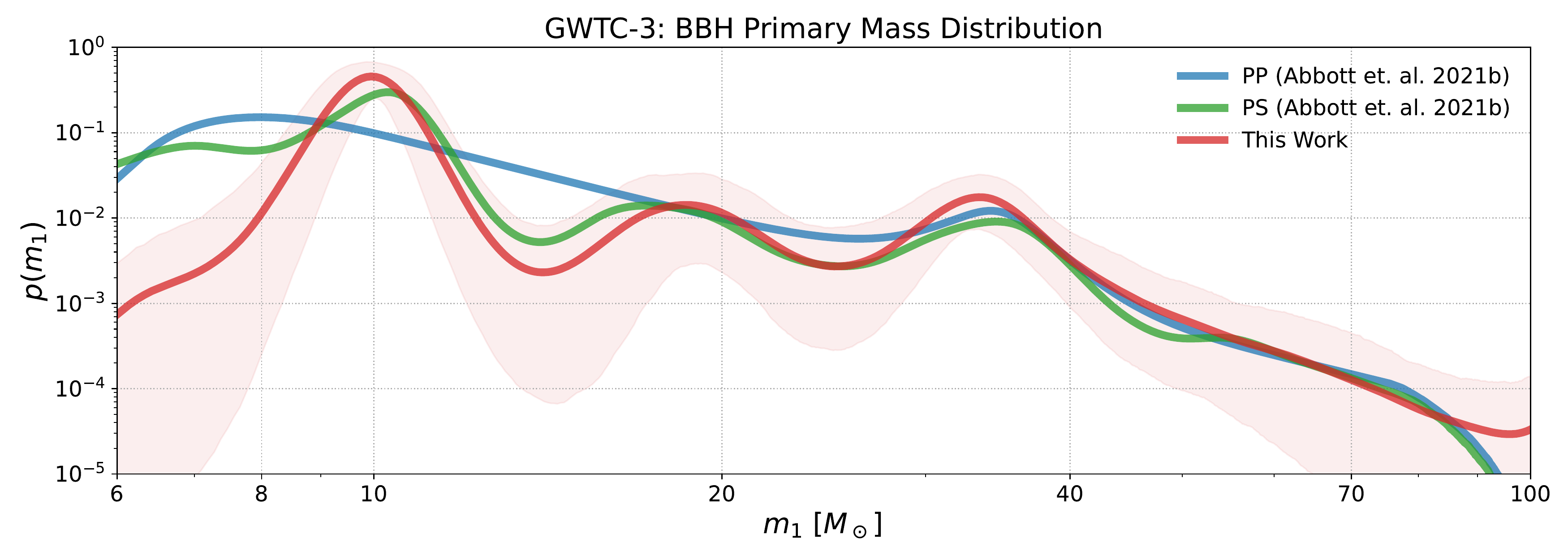}
        \caption{The marginal primary mass distribution inferred with the B-Spline model (red), with 64 knots spaced linearly in $\log m_1$, from 5\msun 
        to 100\msun. The solid line shows the population predictive distribution (PPD), and the shaded region 
        the 90\% credible interval. We show the inferred PPD from the \textsc{PowerlawPeak} (blue) and \textsc{PowerlawSpline} (green) models 
        from the LVK's GWTC-3 population analyses \citep{o3b_astro_dist}. \LinkExplainer}
        \label{fig:mass_distribution}
    \end{centering}
    \script{mass_distribution_plot.py}
\end{figure*}

We use hierarchical Bayesian inference (see Appendix \ref{sec:hierarchical_inference}) to simultaneously infer the astrophysical mass, spin, and redshift distributions of 
binary black holes (BBHs) given a catalog of gravitational wave observations. Following the same cut on the recent GWTC-3 catalog done in the LVK's 
accompanying BBH population analysis, we have 70 possible BBH mergers with false alarm rates less than 1 per year \citep{GWTC3,o3b_astro_dist,GWTC3DATA}. Since it was concluded to be an outlier of the rest of 
the BBH population in both GWTC-2 and GWTC-3, we choose to omit the poorly understood event, GW190814 \citep{190814disc,o3a_pop,o3b_astro_dist,Essick_2022}. This leaves us with 
a catalog of 69 confident BBH mergers, observed over a period of about 2 years, from which we want to infer population properties. 
\NewChange{Following what was done in \citet{o3b_astro_dist}, for events included in GWTC-1 \citep{GWTC1}, we use the published samples that equally weight samples from analyses with the \textsc{IMRPhenomPv2} \citep{1308.3271} and \textsc{SEOBNRv3} \citep{1307.6232,1311.2544} waveforms. For the events from GWTC-2 \citep{gwtc2}, we use samples that equally weight all available analyses using higher-order mode waveforms (\textsc{PrecessingIMRPHM}). Finally, for new events reported in GWTC-2.1 and GWTC-3 \citep{2108.01045,GWTC3}, we use combined samples, equally weighted, from analyses with the \textsc{IMRPhenomXPHM} \citep{2004.06503} and the \textsc{SEOBNRv4PHM} \citep{2004.09442} waveform models.}
Our study provides the first comprehensive 
data-driven investigation, simultaneously inferring all the BBH population distributions (i.e. mass, spin, and redshift), uncovering new insights and corroborating those found with other methods. 
We start with our inference of the mass distribution.

\subsection{Binary Black Hole Masses} \label{sec:mass_dist}

\begin{figure}[ht!]
    \includegraphics[width=\linewidth]{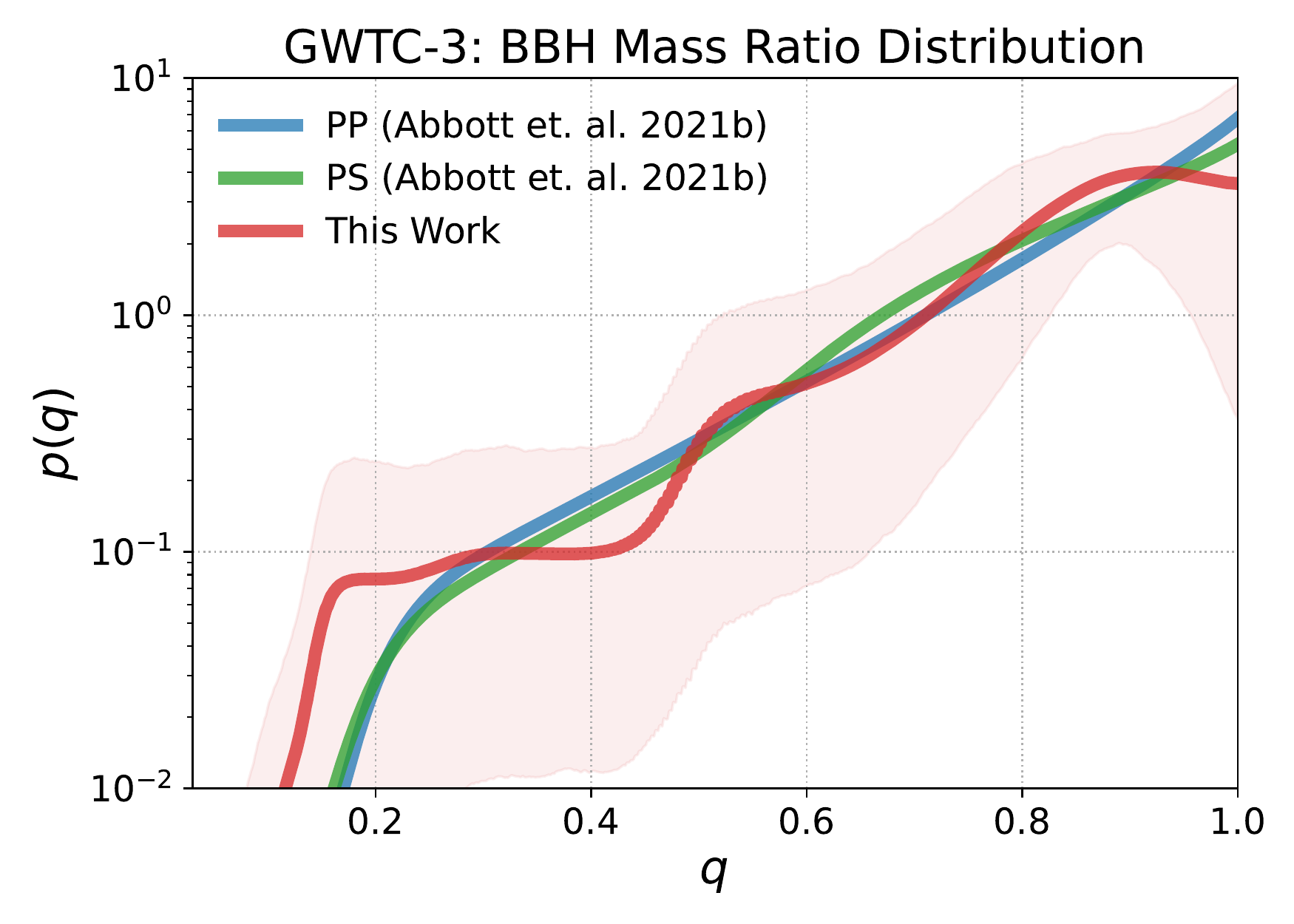}
    \caption{The marginal mass ratio distribution inferred with the B-Spline model (red), with 18 knots spaced linearly in $q$, from 0.05 to 1. The solid line shows the population predictive distribution (PPD), and the shaded region the 90\% credible interval. 
    We show the inferred PPD from the \textsc{PowerlawPeak} (blue) and \textsc{PowerlawSpline} (green) models from the LVK's GWTC-3 population analyses \citep{o3b_astro_dist}. \LinkExplainer}
    \label{fig:q_distribution}
    \script{q_distribution_plot.py}
\end{figure}

Figure \ref{fig:mass_distribution} shows the primary mass distribution inferred with our B-Spline model (red), where we see features consistent with those inferred by the \textsc{PowerlawPeak} and \textsc{PowerlawSpline} mass models \citep{Talbot_2018,o3a_pop,Edelman_2022ApJ,o3b_astro_dist,GWTC3POPDATA}. 
In particular our B-Spline model finds peaks in merger rate density as a function of primary mass at both $\sim10\msun$ and $\sim35\msun$, agreeing with those 
reported using the same dataset in \citet{o3b_astro_dist}.  The B-Spline model finds the same feature at $\sim18\msun$ as the \textsc{PowerlawSpline} model, but remains consistent with the \textsc{PowerlawPeak} model; the mass distribution is more uncertain in this region. For each of these features we find the local maximums occurring at primary masses of \result{$\CIPlusMinus{\macros[MassDistribution][BSpline][peaks][10]}\,\msun$},  
\result{$\CIPlusMinus{\macros[MassDistribution][BSpline][peaks][18]}\,\msun$}, and  
\result{$\CIPlusMinus{\macros[MassDistribution][BSpline][peaks][35]}\,\msun$} all at 90\% credibility.
We find the largest disagreement at low masses, where the power-law-based models show a higher rate below ${\sim}8-9\msun$. This is partly due to the minimum mass hyperparameter (where the power law ``begins'') serving as the minimum allowable primary and secondary masses of the catalog. This leads to inferences of $m_\mathrm{min}$ below the minimum observed primary mass in the catalog, which is $\sim 6.4\msun$, to account for secondary BBH masses lower than that. We choose to fix the minimum black hole mass for both primary and secondaries to $5\msun$, 
similar to the inferred minimum mass in \citet{o3b_astro_dist}. The lack of observations of binaries with low primary mass make rate estimates in this region strongly model dependent, while our flexible model provides an informed upper limit on the rate in this region and given the selection effects and that there are no observations. We could be seeing signs of a decrease in merger rate from a ``lower mass gap'' between neutron star and BH masses, or we could be seeing fluctuations due to low-number statistics \citep{NoPeaksWithoutValleys}. Either way we expect this to be resolved with future catalog updates. We also find no evidence for a sharp fall off in merger rate either following 
the pileup at ${\sim}35\,\msun$ -- expected if such a pileup was due to pulsational pair instability supernovae (PPISNe) -- or where the maximum mass truncation of the power law models 
are inferred. The lack of any high mass truncation, along with the peak at $\sim35\msun$ (significantly lower than expected from PPISNe) may pose challenges for conventional stellar 
evolution theory. This could be hinting at the presence of subpopulations that avoid pair instability supernovae during binary formation, but the confirmation of the existence of such subpopulations 
cannot be determined with the current catalog.

The marginal mass ratio distribution inferred by the B-spline model is shown in figure \ref{fig:q_distribution}.  These results suggest we may be seeing the first signs of departure from a simple power law behavior.  We find a potential signs of a plateau or decrease in the merger rate near equal mass ratios, as well as a broader tail towards unequal mass ratios 
than the power law based models find, although a smooth power law is still consistent with these results given the large uncertainties. Our results also suggest a shallower slope from $q\sim0.3$ to $q\sim0.7$, though uncertainty is larger in this region. The sharp decrease in rate just below 
$q\sim0.5$ is due to the minimum mass ratio truncation defined by $q_\mathrm{min}=\frac{m_\mathrm{min}}{m_1}$. When marginalizing over the primary mass 
distribution with a strong peak at $10\msun$, the mass ratio distribution truncates at $q\sim0.5$: the minimum mass, $5\msun$, 
divided by the most common primary mass, ${\sim}10\msun$. 

\begin{figure}[ht!]
    \begin{centering}    
        \includegraphics[width=\linewidth]{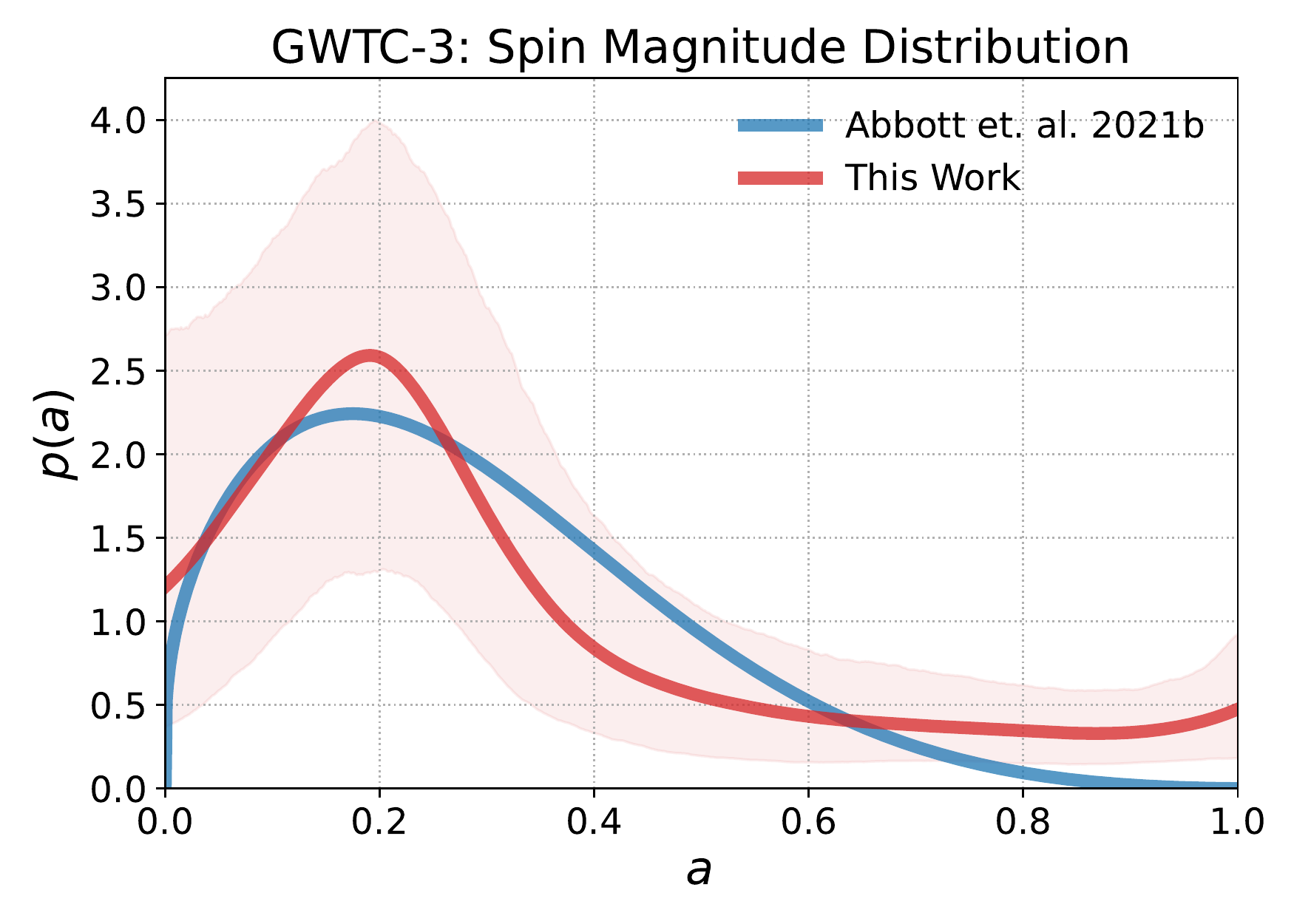}
        \caption{The spin magnitude distribution inferred with the B-Spline model (red) with 16 knots spaced linearly from 0 to 1, assuming the components are IID. The solid line shows the population predictive distribution (PPD), and the shaded region the 90\% credible interval. 
        For comparison, we show the inferred PPD from the \textsc{Default} (blue) model from \citet{o3b_astro_dist}, the LVK's GWTC-3 population analyses. \LinkExplainer}
        \label{fig:iid_spinmag_dist}
    \end{centering}
    \script{plot_iid_spinmag.py}
\end{figure}

\subsection{Binary Black Hole Spins} \label{sec:spin_dist}

\begin{table*}[ht!]
    \centering
    \begin{tabular}{|l|l|l|l|l|l|}
        \hline
        Model & $a_\mathrm{peak}$ & $a_\mathrm{90\%}$ & $\cos{\theta_\mathrm{peak}}$ & $f_{\cos{\theta}<0}$ & $\log_{10}Y$ \\ \hline \hline
        B-Spline IID & $\CIPlusMinus{\macros[BSplineIIDCompSpins][peak_a]}$ & $\CIPlusMinus{\macros[BSplineIIDCompSpins][a_90percentile]}$ & $\CIPlusMinus{\macros[BSplineIIDCompSpins][peakCosTilt]}$ & $\CIPlusMinus{\macros[BSplineIIDCompSpins][negFrac]}$ & $\CIPlusMinus{\macros[BSplineIIDCompSpins][log10gammaFrac]}$ \\ \hline
        B-Spline Ind(primary) & $\CIPlusMinus{\macros[BSplineIndependentCompSpins][peak_a1]}$ & $\CIPlusMinus{\macros[BSplineIndependentCompSpins][a1_90percentile]}$ & $\CIPlusMinus{\macros[BSplineIndependentCompSpins][peakCosTilt1]}$ & $\CIPlusMinus{\macros[BSplineIndependentCompSpins][negFrac1]}$ & $\CIPlusMinus{\macros[BSplineIndependentCompSpins][log10gammaFrac1]}$ \\ \hline
        B-Spline Ind(secondary) & $\CIPlusMinus{\macros[BSplineIndependentCompSpins][peak_a2]}$ & $\CIPlusMinus{\macros[BSplineIndependentCompSpins][a2_90percentile]}$ & $\CIPlusMinus{\macros[BSplineIndependentCompSpins][peakCosTilt2]}$ & $\CIPlusMinus{\macros[BSplineIndependentCompSpins][negFrac2]}$ & $\CIPlusMinus{\macros[BSplineIndependentCompSpins][log10gammaFrac2]}$  \\ \hline
        \textsc{Default} \citep{o3b_astro_dist} & $\CIPlusMinus{\macros[Default][peak_a]}$ & $\CIPlusMinus{\macros[Default][a_90percentile]}$ & $\CIPlusMinus{\macros[Default][peakCosTilt]}$ & $\CIPlusMinus{\macros[Default][negFrac]}$ & $\CIPlusMinus{\macros[Default][log10gammaFrac]}$ \\ \hline
    \end{tabular}
    \caption{Summary of Component Spin distributions inferred both the independent and IID component spin B-Spline models and the \textsc{Default} spin model from \citet{o3b_astro_dist}.}
    \label{tab:compspins}
\end{table*}

\subsubsection{Spin Magnitude}

The \textsc{Default} spin model (used by \citet{o3b_astro_dist}) describes the spin magnitude of both components as identical and independently distributed (IID) non-singular Beta distributions \citep{Talbot_2017,Wysocki_2019}.
The Beta distribution provides a simple 2-parameter model that can produce a wide range of functional forms on the unit interval. However, the constraint that keeps 
the Beta distribution non-singular (i.e. $\alpha>1$ and $\beta>1$) enforces a spin magnitude that always has $p(a_i=0) = 0$. Recent studies have proposed the possible existence of a 
distinct subpopulation of non-spinning or negligibly spinning black holes that can elude discovery with such a model \citep{FullerMa2019,RouletGWTC2Pop,BuildBetterSpinModels,Callister_NoEvidence,GWTC3MonashSpin}. 

\begin{figure}[ht!]
    \begin{centering}
        \includegraphics[width=\linewidth]{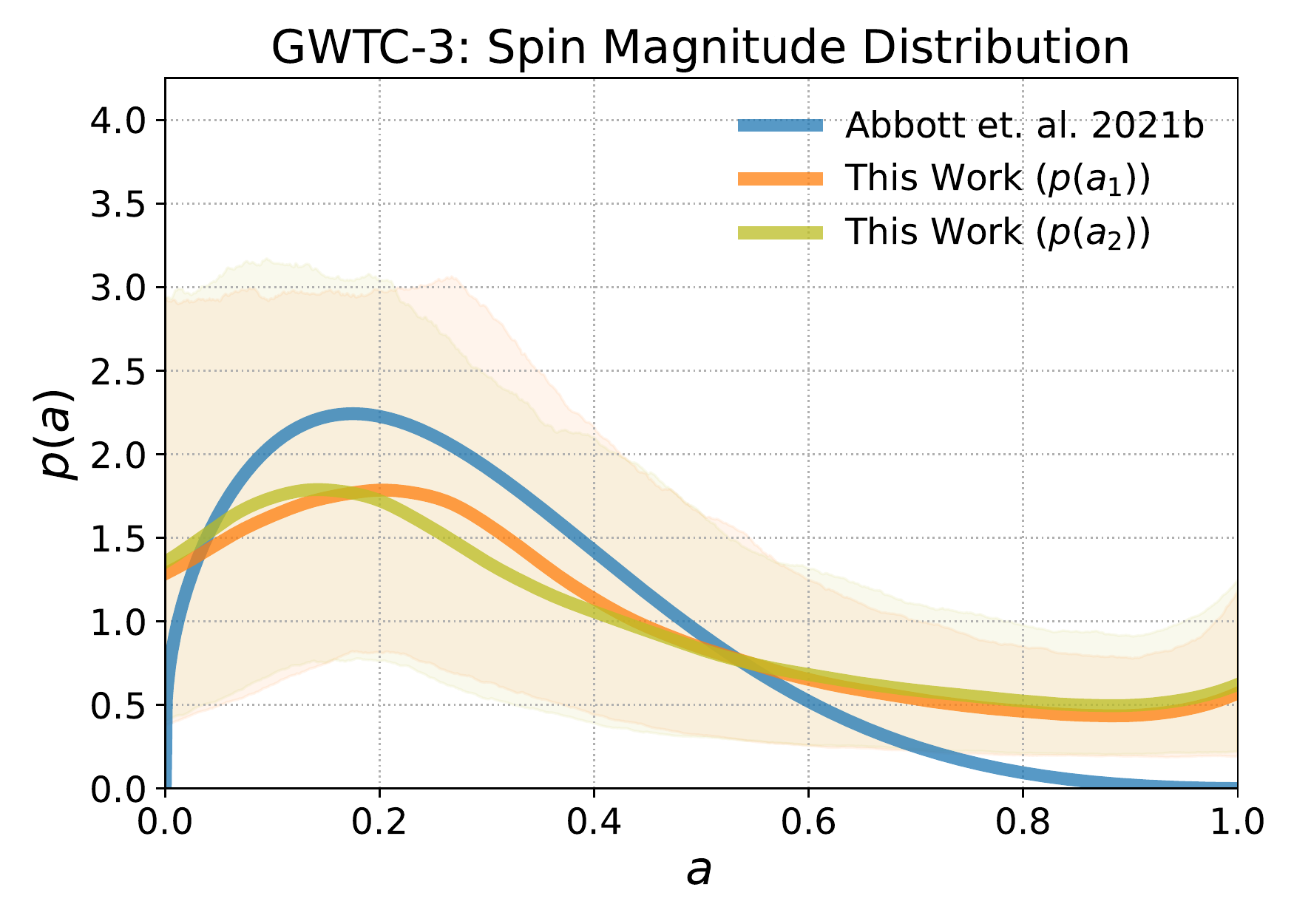}
        \caption{The primary (orange) and secondary (olive) spin magnitude distributions inferred with the B-Spline model
        with 16 knots spaced linearly from 0 to 1. The solid line shows the population predictive distribution (PPD), and the shaded region the 90\% credible interval. 
        For comparison, we show the inferred PPD from the \textsc{Default} (blue) model from \citet{o3b_astro_dist}, the LVK's GWTC-3 population analyses. \LinkExplainer}
        \label{fig:ind_spinmag_dist}
    \end{centering}
    \script{plot_ind_spinmag.py}
\end{figure}

We model the spin magnitude distributions as IID B-Spline distributions. 
Figure \ref{fig:iid_spinmag_dist} shows the inferred spin magnitude distribution with the B-Spline model, compared with the \textsc{Default} model from \citet{o3b_astro_dist}. 
The B-Spline model results are consistent with those using the Beta distribution, peaking near $a\sim0.2$, with 90\% of BBH spins below 
\result{$\CIPlusMinus{\macros[BSplineIIDCompSpins][a_90percentile]}$} at 90\% credibility. The B-Spline model does not impose vanishing support at the extremal values like the Beta distribution, 
allowing it to probe the zero-spin question. We find broad support, with large variance, for non-zero probabilities at $a_i=0$, but cannot confidently determine the presence of 
a significant non-spinning subpopulation, corroborating similar recent conclusions \citep{BuildBetterSpinModels,Callister_NoEvidence,GWTC3MonashSpin,Mould2022}. 
We repeat the same analysis with independent B-Spline distributions for each spin magnitude component. In figure \ref{fig:ind_spinmag_dist} 
we show the inferred primary (orange), and secondary (olive) spin magnitude distributions inferred when relaxing the IID assumption. We find no signs that the spin magnitude distributions are not IID but that the primary spin magnitude distribution peaks slightly higher, at $a\sim0.25$, than the IID B-Spline model in figure \ref{fig:iid_spinmag_dist}, but with similar support at near vanishing spins. The secondary spin magnitude distribution is more uncertain due to the 
higher measurement uncertainty when inferring the secondary spins of BBH systems \citep{1403.0129,1611.01122}. The secondary distribution also peaks at smaller spin magnitudes of $a\sim0.15$, showing potentially rates at $a\sim0$ than the primary distribution or B-Spline IID spin magnitude distribution in figure \ref{fig:iid_spinmag_dist}, though uncertainties are large. While the distributions are broadly consistent, we could be seeing signs that component spin magnitude distributions are uniquely distributed, which can be produced through mass-ratio reversal in isolated binary evolution \citep{Mould2022}.

\subsubsection{Spin Orientation}

The \textsc{Default} spin model (used in \citet{o3a_pop, o3b_astro_dist}) also assumes the spin orientation of both components are identical and independently distributed (IID), with a mixture model over an
aligned and an isotropic component. The aligned component is modeled with a truncated Gaussian distribution with mean at $\cos{\theta}=1$ and variance a free 
hyperparameter to be fit \citep{Talbot_2017,Wysocki_2019,o3a_pop,o3b_astro_dist}. This provides a simple 2-parameter model motivated by simple distributions expected from the two main formation scenario families, allowing 
for a straightforward interpretation of results. One possible limitation however, is that by construction this distribution is forced to peak at perfectly aligned spins, 
i.e. $\cos{\theta}=1$. While this may be a reasonable assumption, \citet{spinitasyoulike} recently extended the model space of parametric descriptions 
used to model the spin orientation distribution and found considerable evidence that the distribution peaks away from $\cos{\theta}=1$. Again, this provides a clear 
use-case where data-driven models can help us understand the population.

\begin{figure}[ht!]
    \begin{centering}
        \includegraphics[width=\linewidth]{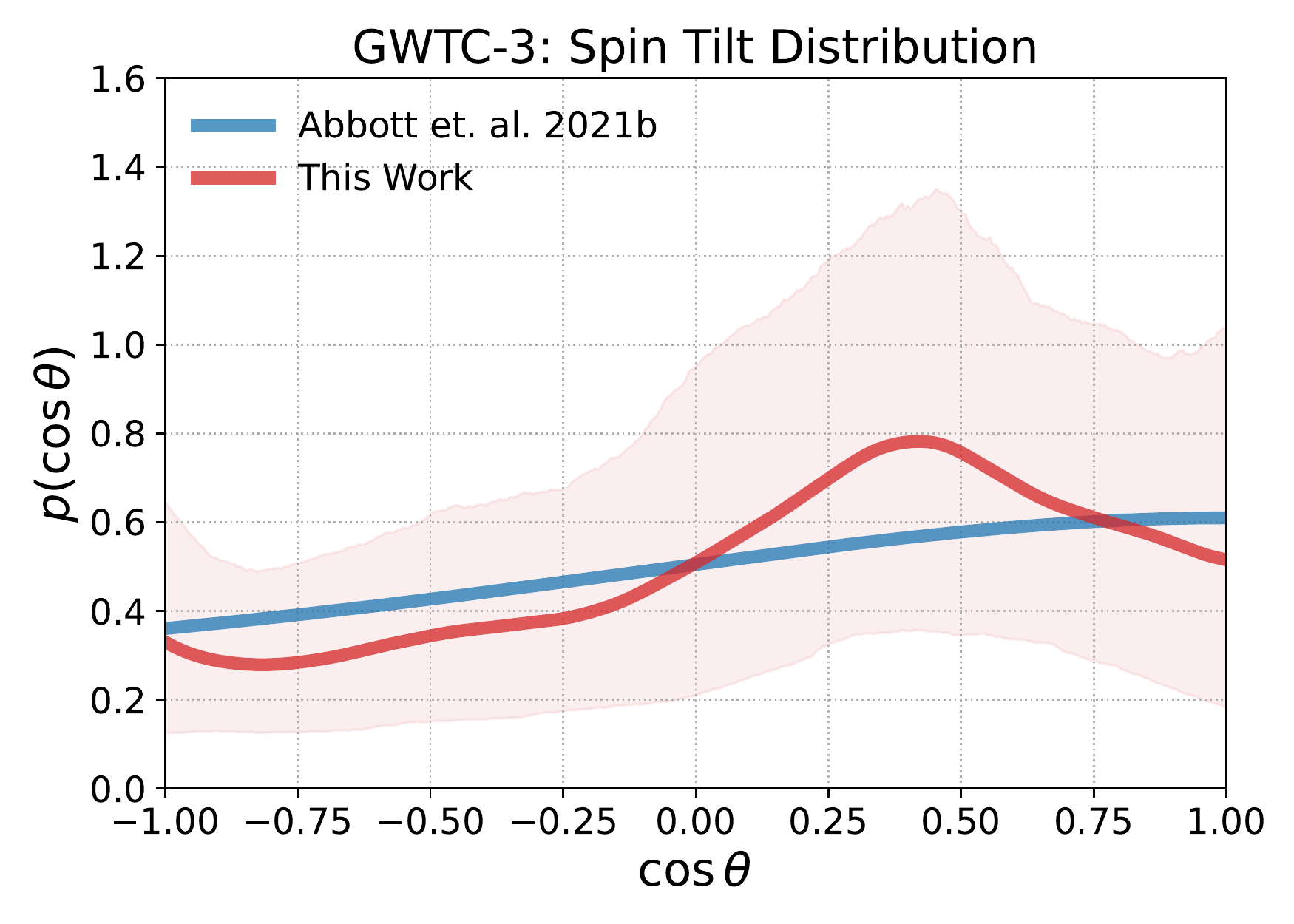}
        \caption{The spin orientation distribution inferred with the B-Spline model (red) with 16 knots spaced linearly from -1 to 1, and assuming the components are IID. The solid line shows the population predictive distribution (PPD), and the shaded region the 90\% credible interval. 
        For comparison, we show the inferred PPD from the \textsc{Default} (blue) model from \citet{o3b_astro_dist}, the LVK's GWTC-3 population analyses. \LinkExplainer}
        \label{fig:iid_spintilt_dist}
    \end{centering}
    \script{plot_iid_spintilt.py}
\end{figure}

\NewChange{Figure \ref{fig:iid_spintilt_dist} shows the inferred spin orientation distribution with the IID spin B-Spline model}, compared with the \textsc{Default} model from \citet{o3b_astro_dist}. 
The B-Spline inferences have large uncertainties but start to show the same features as found and discussed in \citet{spinitasyoulike}. 
We find a distribution that instead of intrinsically peaking at $\cos{\theta}=1$, is found to peak at: $\cos{\theta}=$\result{$\CIPlusMinus{\macros[BSplineIIDCompSpins][peakCosTilt]}$}, at 
90\% credibility. We find less, but still considerable support for misaligned spins (i.e. $\cos{\theta}<0$), consistent with other recent studies \citep{o3a_pop,o3b_astro_dist,Callister_NoEvidence}. Specifically we 
find that the fraction of misaligned systems is $f_{\cos{\theta}<0}=$\result{$\CIPlusMinus{\macros[BSplineIIDCompSpins][negFrac]}$}, compared to 
$f_{\cos{\theta}<0}=$\result{$\CIPlusMinus{\macros[Default][negFrac]}$} with the \textsc{Default} model from \citet{o3b_astro_dist}. This implies 
the presence of an isotropic component as expected by dynamical formation channels, albeit less than with the \textsc{Default} model. To quantify the 
amount of isotropy in the tilt distribution we calculate $\log_{10}Y$, where $Y$ is the ratio of nearly aligned tilts to nearly anti-aligned, 
introduced in \citet{spinitasyoulike} and defined as:
\begin{equation}
    Y \equiv \frac{\int_{0.9}^{1.0} d\cos{\theta} p(\cos{\theta})}{\int_{-1.0}^{-0.9} d\cos{\theta} p(\cos{\theta})}.
\end{equation}
The log this quantity, $\log_{10}Y$, is 0 for tilt distribution that is purely isotropic, negative when anti-aligned values are favored, 
and positive when aligned tilts are favored. We find a $\log_{10}Y=$\result{$\CIPlusMinus{\macros[BSplineIIDCompSpins][log10gammaFrac]}$}, exhibiting a slight preference for aligned tilts.  

\begin{figure}[ht!]
    \begin{centering}
        \includegraphics[width=\linewidth]{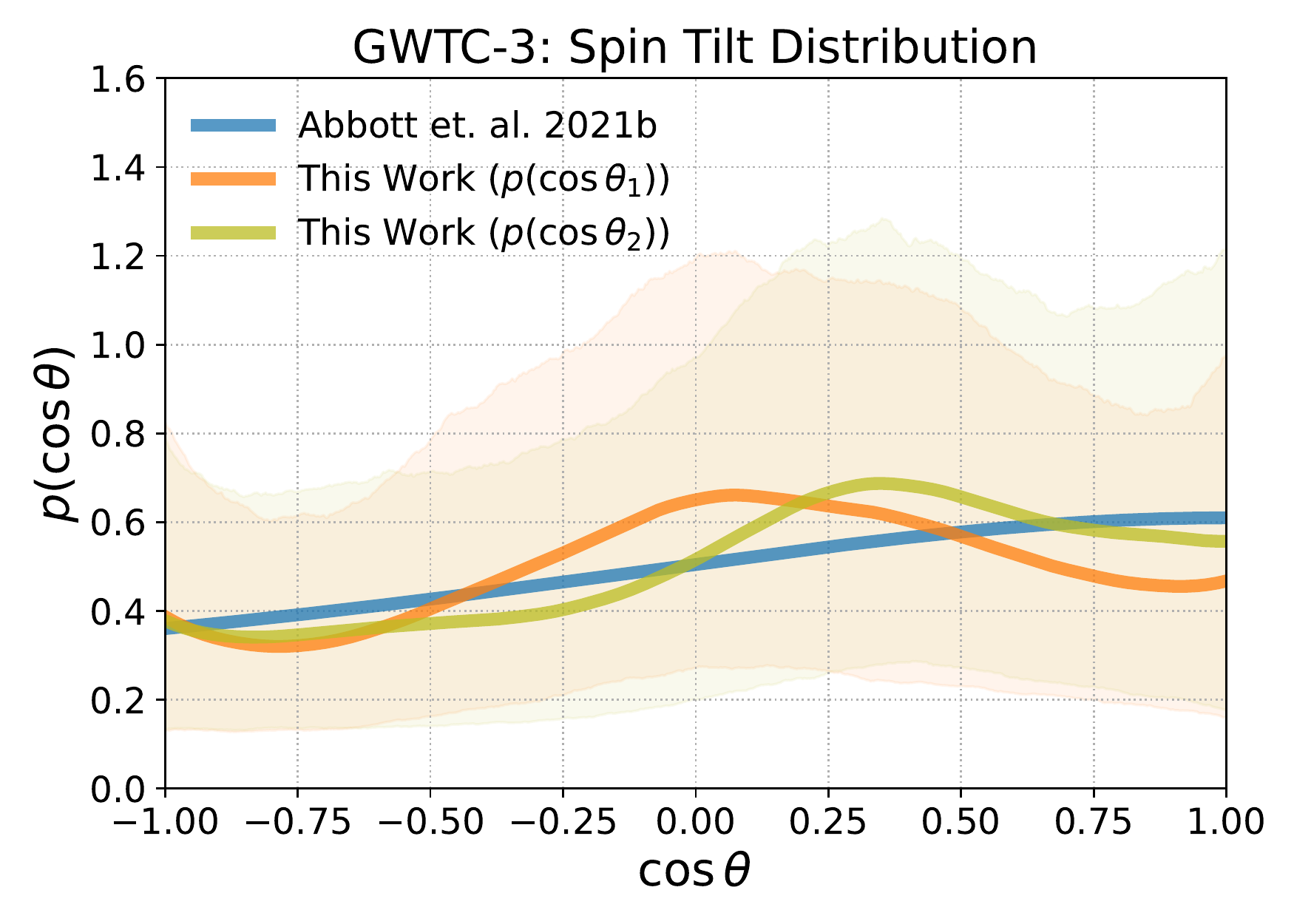}
        \caption{The primary (orange) and secondary (olive) spin orientation distributions inferred with the B-Spline model with 16 knots spaced linearly from -1 to 1. The solid line shows the population predictive distribution (PPD), and the shaded region the 90\% credible interval. 
        For comparison, we show the inferred PPD from the \textsc{Default} (blue) model from \citet{o3b_astro_dist}, the LVK's GWTC-3 population analyses. \LinkExplainer}
        \label{fig:ind_spintilt_dist}
    \end{centering}
    \script{plot_ind_spintilt.py}
\end{figure}

We also model each component's orientation distribution with an independent B-Spline model as done above, and show the inferred 
primary (orange), and secondary (olive) distributions in figure \ref{fig:ind_spintilt_dist}. The orientation distributions are broadly
consistent with each other and the \textsc{Default} model's PPD given the wide credible intervals. We find the two distributions to peak at: $\cos{\theta_1}=$\result{$\CIPlusMinus{\macros[BSplineIndependentCompSpins][peakCosTilt1]}$} 
and $\cos{\theta_2}=$\result{$\CIPlusMinus{\macros[BSplineIndependentCompSpins][peakCosTilt2]}$}, showing that the primary distribution peak is inferred further away 
from the assumed $\cos{\theta}=1$ with the \textsc{Default} model. There is also significant (albeit uncertain) evidence of spin misalignment in each distribution, finding 
the fraction of misaligned primary and secondary components as: $f_{\cos{\theta_1}<0}=$\result{$\CIPlusMinus{\macros[BSplineIndependentCompSpins][negFrac1]}$} and 
$f_{\cos{\theta_2}<0}=$\result{$\CIPlusMinus{\macros[BSplineIndependentCompSpins][negFrac2]}$}. We again calculate $\log_{10}Y$  
for each component distribution and find: $\log_{10}Y_1=$\result{$\CIPlusMinus{\macros[BSplineIndependentCompSpins][log10gammaFrac1]}$} and 
$\log_{10}Y_2=$\result{$\CIPlusMinus{\macros[BSplineIndependentCompSpins][log10gammaFrac2]}$}.

\begin{table*}[ht!]
    \centering
    \begin{tabular}{|l|l|l|l|l|l|}
        \hline
        Model & $\chi_\mathrm{eff,peak}$ & $f_{\chi_\mathrm{eff}<0}$ & $f_{\chi_\mathrm{eff}<-0.3}$ & $f_\mathrm{dyn}$ & $f_\mathrm{HM}$ \\ \hline \hline
        B-Spline IID & $\CIPlusMinus{\macros[ChiEffective][iid][PeakChiEff]}$ & $\CIPlusMinus{\macros[ChiEffective][iid][FracBelow0]}$ & $\CIPlusMinus{\macros[ChiEffective][iid][FracBelowNeg0p3]}$ & $\CIPlusMinus{\macros[ChiEffective][iid][frac_dyn]}$ & $\CIPlusMinus{\macros[ChiEffective][iid][frac_hm]}$ \\ \hline
        B-Spline Ind & $\CIPlusMinus{\macros[ChiEffective][ind][PeakChiEff]}$ & $\CIPlusMinus{\macros[ChiEffective][ind][FracBelow0]}$ & $\CIPlusMinus{\macros[ChiEffective][ind][FracBelowNeg0p3]}$ & $\CIPlusMinus{\macros[ChiEffective][ind][frac_dyn]}$ & $\CIPlusMinus{\macros[ChiEffective][ind][frac_hm]}$ \\ \hline
        \textsc{Default} \citep{o3b_astro_dist} & $\CIPlusMinus{\macros[ChiEffective][default][PeakChiEff]}$ & $\CIPlusMinus{\macros[ChiEffective][default][FracBelow0]}$ & $\CIPlusMinus{\macros[ChiEffective][default][FracBelowNeg0p3]}$ & $\CIPlusMinus{\macros[ChiEffective][default][frac_dyn]}$ & $\CIPlusMinus{\macros[ChiEffective][default][frac_hm]}$ \\ \hline
        \textsc{Gaussian} \citep{o3b_astro_dist}  & $\CIPlusMinus{\macros[ChiEffective][gaussian][PeakChiEff]}$ & $\CIPlusMinus{\macros[ChiEffective][gaussian][FracBelow0]}$ & $\CIPlusMinus{\macros[ChiEffective][gaussian][FracBelowNeg0p3]}$ & $\CIPlusMinus{\macros[ChiEffective][gaussian][frac_dyn]}$ & $\CIPlusMinus{\macros[ChiEffective][gaussian][frac_hm]}$ \\ \hline
    \end{tabular}
    \caption{Summary of the effective spin distributions inferred with the B-Spline model variations, along with the \textsc{Default} and \textsc{Gaussian} models from \citet{o3b_astro_dist}.}
    \label{tab:chieff}
\end{table*}

\subsection{The Effective Spin Dimension}

\begin{figure*}[ht!]
    \begin{centering}
        \includegraphics[width=\linewidth]{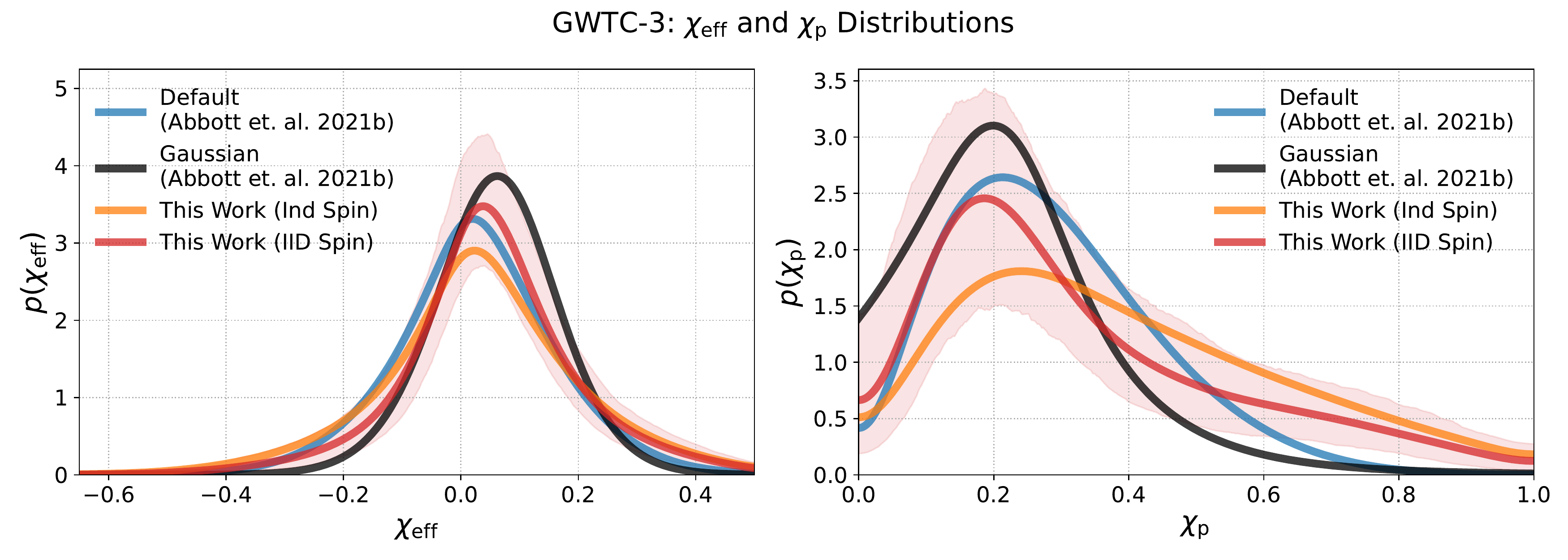}
        \caption{The effective (left) and precessing (right) spin distributions inferred with the B-Spline IID spin model (red). The solid line shows the population predictive distribution (PPD), and the shaded region the 90\% credible interval. We show the inferred PPDs from the independent component spin B-Spline model (purple), and both the \textsc{Default} (blue) model and the 
        Gaussian (green) model from \citet{o3b_astro_dist}, the LVK's GWTC-3 population analyses. \LinkExplainer}
        \label{fig:eff_dist}
    \end{centering}
    \script{plot_effspin.py}
\end{figure*}

\NewChange{While the component spin magnitudes and tilts are more directly tied to formation physics, they are typically poorly measured.  The best-measured spin quantity, which enters at the highest post-Newtownian order, is the effective spin: $\chi_\mathrm{eff} = \frac{a_1\cos{\theta_1} + qa_2\cos{\theta_2}}{1+q}$. There is additionally an effective precessing spin parameter, $\chi_\mathrm{p} = \mathrm{max}\big[ a_1 \sin{\theta_1}, \frac{3+4q}{4+3q} q a_2 \sin{\theta_2} \big]$, that quantifies the amount of spin precession given the systems mass ratio and component spin magnitudes and orientation.}
Figure \ref{fig:eff_dist} shows the inferred effective spin and precessing spin distributions with the two versions of our B-Spline models (red and purple), 
along with results on the \textsc{Default} \citep{Talbot_2017} and \textsc{Gaussian} \citep{Miller2020} models from \citet{o3b_astro_dist}. We find considerable agreement 
among the effective spin distributions, but the more flexible B-Spline models in component spins more closely resemble results from the \textsc{Default} model, also using the component spins. The B-Spline model finds very similar shapes to the other models, with a single peak centered at 
$\chi_\mathrm{eff}=$\result{$\CIPlusMinus{\macros[ChiEffective][iid][PeakChiEff]}$}, compared to 
$\chi_\mathrm{eff}=$\result{$\CIPlusMinus{\macros[ChiEffective][default][PeakChiEff]}$} with the \textsc{Default} model and 
$\chi_\mathrm{eff}=$\result{$\CIPlusMinus{\macros[ChiEffective][gaussian][PeakChiEff]}$} with the \textsc{Gaussian} $\chi_\mathrm{eff}$ models from \citet{o3b_astro_dist}. 
As for spin misalignment, we calculate the fraction of systems with effective spins that are misaligned (i.e. $\chi_\mathrm{eff}<0$) and find similar 
agreement with previous work \citep{o3a_pop,o3b_astro_dist,Callister_NoEvidence}. We find for the B-Spline model 
$f_{\chi_\mathrm{eff}<0}=$\result{$\CIPlusMinus{\macros[ChiEffective][iid][FracBelow0]}$}, compared to 
$f_{\chi_\mathrm{eff}<0}=$\result{$\CIPlusMinus{\macros[ChiEffective][default][FracBelow0]}$} and $f_{\chi_\mathrm{eff}<0}=$\result{$\CIPlusMinus{\macros[ChiEffective][gaussian][FracBelow0]}$} 
with the \textsc{Default} and \textsc{Gaussian} models from \citet{o3b_astro_dist}. \NewChange{The precessing spin distributions inferred with the B-Spline models exhibit a similar shape to the \textsc{Default} model, but with a much fatter tail towards highly precessing systems, driven by the extra support for highly spinning components seen in figures \ref{fig:iid_spinmag_dist} and \ref{fig:iid_spinmag_dist}.}

\begin{figure}[ht!]
    \includegraphics[width=\linewidth]{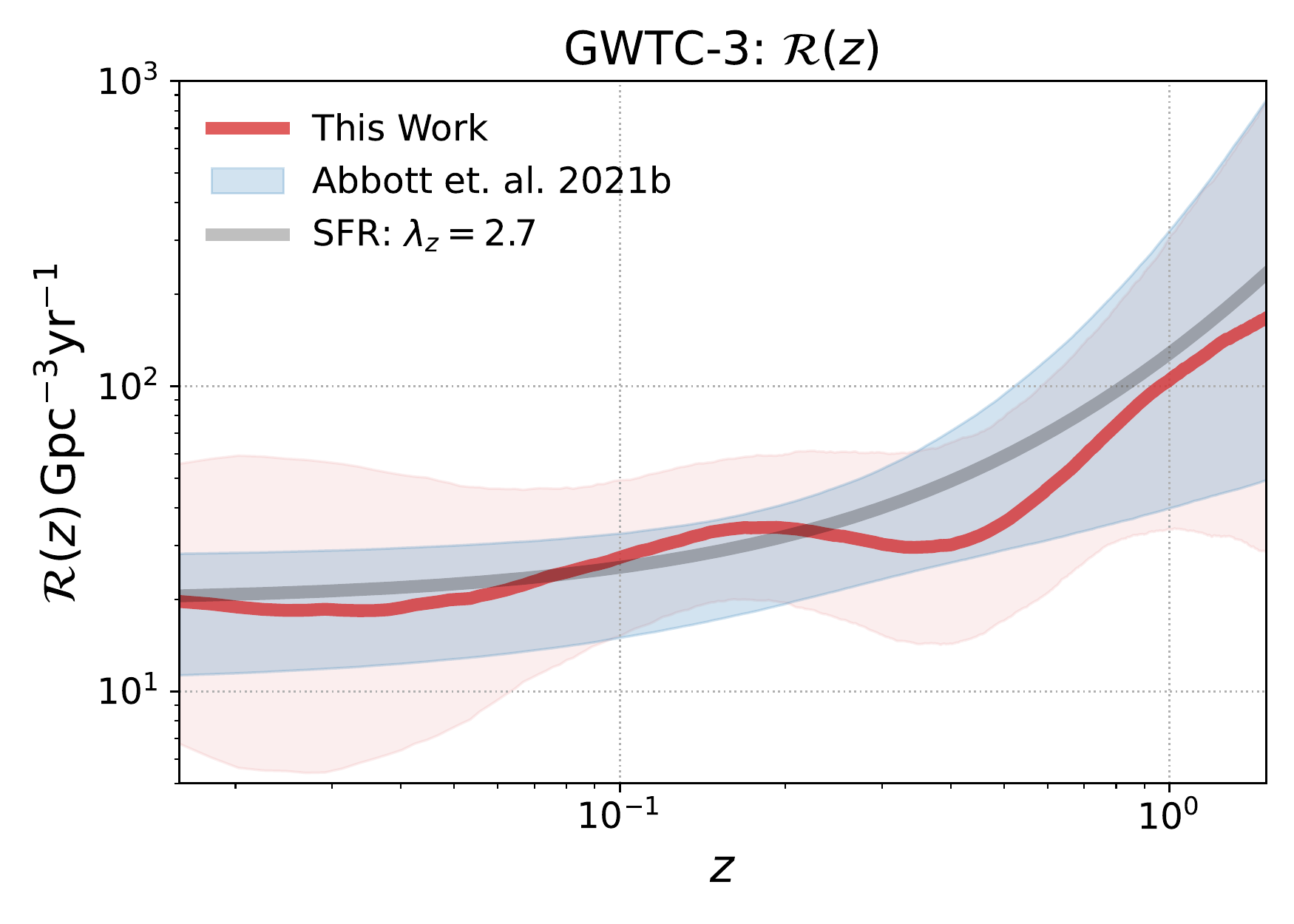}
    \caption{The BBH merger rate as a function of redshift. We show the B-Spline model (red) with 16 knots spaced linearly in $\log(z)$, 
    from the minimum to the maximum observed redshifts. The solid line shows the population predictive distribution (PPD), and the shaded region 
    the 90\% credible interval. We show the inferred 90\% credible interval from the \textsc{PowerlawRedshift} model 
    from the LVK's GWTC-3 population analyses in blue and a power law with exponent of 2.7 in gray, representing the expected star formation rate 
    \citep{Madau_2014, o3b_astro_dist}. \LinkExplainer}
    \label{fig:rofz}
    \script{rate_vs_z_plot.py}
\end{figure}

\subsection{Merger Rate Evolution with Redshift} \label{sec:redshift}

\begin{figure}[ht!]
    \includegraphics[width=\linewidth]{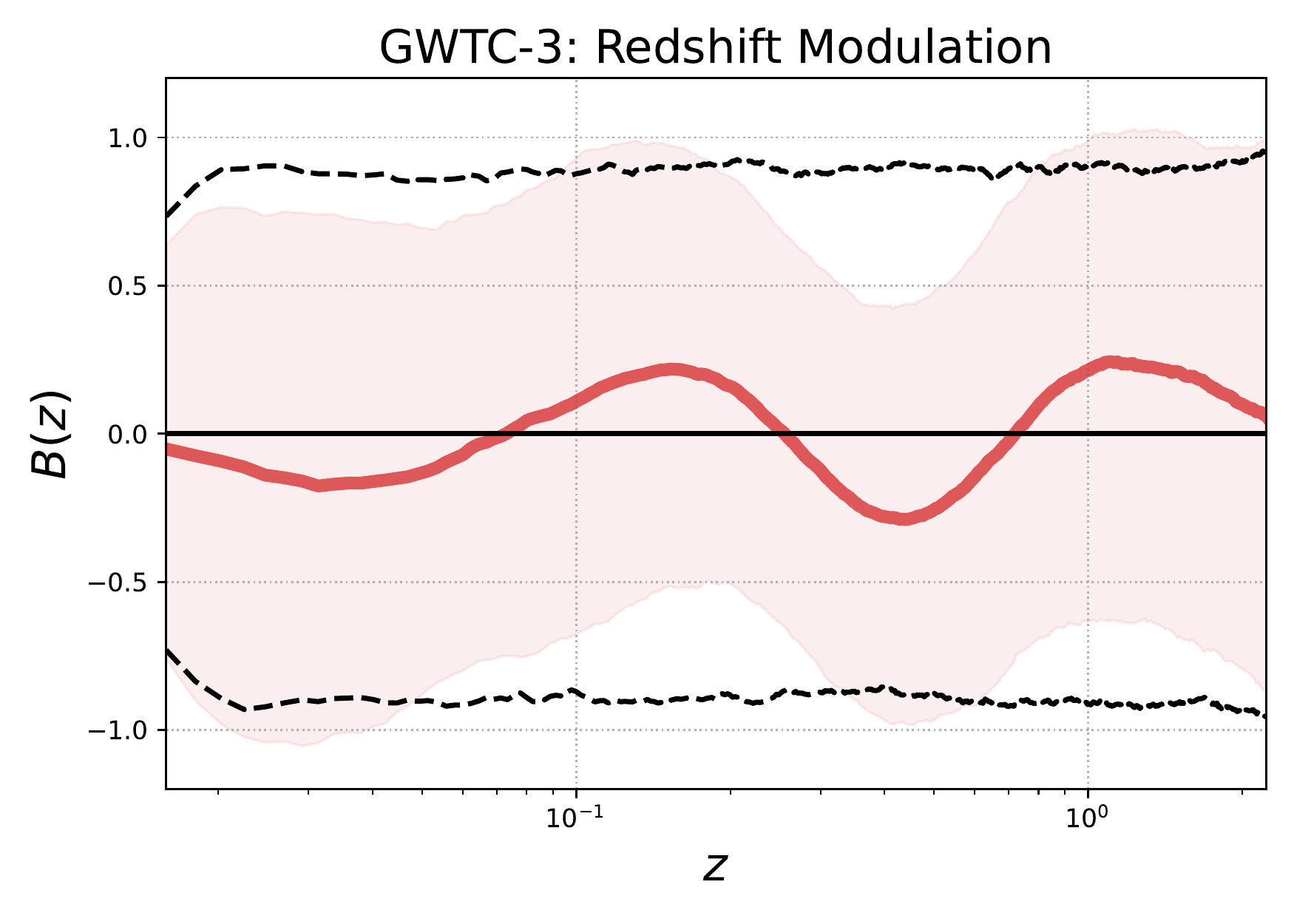}
    \caption{The B-spline modulation to the underlying power law in redshift, (red). The solid line shows the population predictive distribution (PPD),
    and the shaded region the 90\% credible interval. We show the 90\% credible interval of the prior predictive distribution in dashed black lines. \LinkExplainer}
    \label{fig:z_modulation}
    \script{z_modulation_plot.py}
\end{figure}

Recent analysis of the GWTC-3 BBH population has shown evidence for an increasing merger rate with redshift, nearly ruling out a merger rate that is 
constant with co-moving volume \citep{Fishbach_2018redshift,o3b_astro_dist}. When extending the power law form of the previously used model to have a modulation 
that we model with B-Splines, the merger rate as a function of redshift in figure \ref{fig:rofz} shows mild support for features departing from the underlying power law. 
In particular, we see a small increase in merger rate from $z\sim0.09$ to $z\sim0.2$ (where we best constrain the rate), followed by a plateau in the rate from $z\sim0.2$ to $z\sim0.4$. 
At larger redshifts, where we begin to have sparse observations, we see no sign of departure from the power-law as the rate continues to increase with redshift. 
The underlying power-law slope of our B-spline modulated model is consistent with the GWTC-3 results with the underlying model by itself: the \textsc{PowerlawRedshift} model 
found $\lambda_z = $\result{$\CIPlusMinus{\macros[PLPeak][lamb]}$} when inferred with the \textsc{PowerlawPeak} mass, and \textsc{Default} spin models. Our more 
flexible model infers a power law slope of $\lambda_z = $\result{$\CIPlusMinus{\macros[BSplineIIDCompSpins][lamb]}$}. We show the basis spline modulations or departure 
from the power law in \ref{fig:z_modulation} compared to the prior -- showing where we cannot constrain any significant deviations from the simpler parametric power law model. 
The extra freedom of our model does inflate the uncertainty in its rate estimates, especially at $z\sim0$ where there are not any observations in the catalog. 
We find a local ($z=0$) merger rate of $\mathcal{R}_0 = $\result{$\CIPlusMinus{\macros[BSplineIIDCompSpins][local_rate]}\,\mathrm{Gpc}^{-3}\mathrm{yr}^{-1}$} using the B-Spline modulation model 
which compares to $\mathcal{R}_0 = $\result{$\CIPlusMinus{\macros[PLPeak][local_rate]}\,\mathrm{Gpc}^{-3}\mathrm{yr}^{-1}$} for the GWTC-3 result. 

%% file: conclusion.tex
\section{Astrophysical Implications}\label{sec:astrodiscussion}

The collective distribution of BBH source properties provides a useful probe of the complex and uncertain astrophysics that govern their 
formation and evolution until merger \citep{Rodriguez_2016,Farr2017Nature,Zevin_2017}. Our analyses with the newly constructed B-spline models uncover hints new features in the population (e.g., in mass ratio and redshift) and corroborates important 
conclusions of recent work, and provides a robust data-driven framework for future population studies. 


The results presented in section \ref{sec:mass_dist} illustrate a wider mass distribution than inferred with power-law based models in \citet{o3b_astro_dist}, and a suppressed merger rate 
at low primary masses (i.e. $\leq8\msun$), showing possible signs of binary selection effects or the purported low mass gap between neutron stars and black holes \citep{Fishbach_2020_mm,Farah_2022,NoPeaksWithoutValleys}. While isolated formation is able to predict the $10\msun$ peak \citep{Antonini_2020}, cluster and dynamical formation scenarios struggle to predict 
a peak in the BH mass distribution less than $15-20\msun$ \citep{Hong_2018, Rodriguez_2019}. Globular cluster formation is expected to produce more top-heavy mass distributions than isolated 
and recent studies have shown suppressed BBH merger rates at lower ($m\leq15\msun$) masses when compared to predictions from the isolated channel \citep{Rodriguez_2015, Rodriguez_2019, BaveraMassTransfer,Belczynski_2016}. 
BBHs that form near active galactic nuclei (AGN) can preferentially produce higher mass black holes \citep{FordAGN, Tagawa_2021, Yang_2019}. 
We do not find any evidence for a truncation or rapid decline in the merger rate as a function of mass, that stellar evolution theory predicts due to pair-instability supernovae (PISNe) \citep{Heger_2002,PISN_Woosley,Heger_2003,Spera_2017,Stevenson_2019}. The original motivation for the peak in the \textsc{PowerlawPeak} model \citep{Talbot_2018} was to represent a possible ``pileup'' of 
masses just before such truncation, since massive stars just light enough to avoid PISN will shed large amounts of mass in a series of ``pulses'' before collapsing to BHs in a process called 
pulsational pair-instability supernova (PPISN) \citep{Woosley_2017,Woosley_2019,Farmer_2019}. While the predictions of the mass scale where pair-instability kicks in are uncertain and depend on poorly understood physics 
like nuclear reaction rates of carbon and oxygen in the core of stars, models have a hard time producing this peak lower than $m\sim40\msun$ \citep{Belczynski_2016,Marchant_2019,Renzo_2020,Farmer_2019,Farmer_2020}. The lack of a truncation could 
point towards a higher prevalence of dynamical processes that can produce black holes in mass ranges stellar collapse cannot, such as hierarchical mergers of BHs \citep{Fishbach_2017,Doctor_2020,Kimball_genealogy,kimball2020evidence,doctor2021black,Fishbach_2022}, 
very low metallicity population III stars \citep{Belczynski_2020,Farrell_2020}, new beyond-standard-model physics\citep{Croon_newphysics,Sakstein_2020}, or black hole accretion of BHs in gaseous environments such as AGNs \citep{Secunda_2020,McKernan_2020,cruzosorio2021gw190521}. 

Our constraints on the mass ratio distribution are not yet precise enough to claim definitive departures from power law behavior, but do suggest possible plateaus in the rate at several mass ratios, including equal mass.  These features should sharpen (or resolve) with future updates to the catalog.

Section \ref{sec:spin_dist} focused on inferences of the spin distributions of black holes, observing evidence of spin misalignment, spin anti-alignment, and suppressed support 
for exactly aligned systems. These point towards a significant contribution to the population from dynamical formation processes, agreeing with 
conclusions drawn about the mass distribution inference of section \ref{sec:mass_dist}. While field formation is expected to produce systems with preferentially 
aligned spins due to tidal interactions, observational evidence suggests that tides may not be able to re-align spins in all systems as some 
isolated population models assume. Additionally, because of uncertain knowledge of supernovae kicks, isolated formation can produce systems with negative but small effective spins. 
Consistent with recent studies we report an effective spin distribution that is not symmetric about zero, disfavoring a scenario in which all BBHs are formed dynamically \citep{o3a_pop,o3b_astro_dist,Callister_NoEvidence}. 
Following the rules in \citet{Fishbach_2022}, 
we place conservative upper bounds on the fraction of hierarchical mergers $f_\mathrm{HM}$ and fraction of dynamically formed BBHs, $f_\mathrm{dyn}$ with 
the B-spline $\chi_\mathrm{eff}$ model constraining $f_\mathrm{HM} < $\result{$\macros[ChiEffective][iid][frac_hm][10th percentile]$} 
and $f_\mathrm{dyn} < $\result{$\macros[ChiEffective][iid][frac_dyn][10th percentile]$} at 90\% credibility. This is consistent with the 90\% credible interval found from the GWTC-2 analysis, $0.25\leq f_\mathrm{dyn} \leq 0.93$ \citep{o3a_pop}. 

Finally, section \ref{sec:redshift} shows potentially interesting evolution of the BBH merger rate with redshift.  Though uncertainties are still large, we may be seeing the first signs of departure from following the star formation rate, which could help in distinguishing different subpopulations should they exist \citep{van_Son_2022}.  Again, we expect these features to be resolved with future catalogs.

\section{Conclusions}\label{sec:conclusion}

Non-parametric and data-driven statistical modeling methods have been put to use with great success across the ever-growing field of gravitational 
waves \citep{B_Farr_etal_2014,Littenberg_2015,Mandel_2016,Edwards_2018,Doctor_GPR,Edelman_2021,Vitale_2021,Tiwari_2021_a,Tiwari_2021_b,Edelman_2022ApJ,Tiwari_2022ApJ,ModelExploration_MaxPopLike}. 
We presented a case study exploring how basis splines make for an especially powerful and efficient data driven method of characterizing the binary black hole population observed 
with gravitational waves, along with the associated open source software \href{https://git.ligo.org/bruce.edelman/gwinferno}{GWInferno}, that implements the models 
described in this paper and performs hierarchical Bayesian inference with \textsc{NumPyro} and \textsc{Jax} \citep{pyro,numpyro,jax}. 
Our study paves the way as the first completely non-parametric compact object population study, employing data driven models for each of the hierarchically 
modeled population distributions. A complete understanding of the population properties of compact objects will help to advance poorly understood areas of stellar and 
nuclear astrophysics and provide a novel independent cosmological probe. With the coming influx of new data with the LVK's next observing run, development of model-agnostic methods, such as the one we proposed here, will become necessary to efficiently make
sense of the vast amounts of data and to extract as much information as possible from the population. 

%% file: appendicies.tex
\appendix
\section{Basis Splines} \label{sec:basis_splines}

A common non-parametric method used in many statistical applications is basis splines. A spline function of order $k$, 
is a piece-wise polynomial of order $k$ polynomials stitched together from defined ``knot'' locations across the domain. 
They provide a useful and cheap way to interpolate generically smooth functions from a finite sampling of ``knot'' heights. 
Basis splines of order $k$ are a set of order $k$ polynomials that form a complete basis for any spline function of order $k$. 
Therefore, given an array of knot locations, $\mathbf{t}$ or knot vector, there exists a single unique linear combination of basis splines 
for every possible spline function interpolated from $\mathbf{t}$. To construct a basis of $n$ components and knots, $t_0$, $t_1$,...,$t_{i+k}$, 
we use the Cox-de Boor recursion formula \citep{deBoor78,monotone_regression_splines}. The recursion starts with the $k=0$ (constant) case and recursively constructs 
the basis components of higher orders. The base case and recursion relation that generates this particular basis are defined as:

\begin{equation}
    B_{i,0}(x | \mathbf{t}) = 
    \begin{cases}
        1, & \text{if } t_i \leq x < t_{i+1} \\
        0, & \text{otherwise}
    \end{cases}
\end{equation}

\begin{equation}
    B_{i,k+1}(x | \mathbf{t}) = \omega_{i,k}(x | \mathbf{t})B_{i,k}(x | \mathbf{t}) + \big[1-\omega_{i+1,k}(x | \mathbf{t})\big] B_{i+1,k}(x | \mathbf{t})
\end{equation}

\begin{equation}
    \omega_{i,k}(x | \mathbf{t}) =
    \begin{cases}
        \frac{x-t_i}{t_{i+k}-t_i}, & t_{i+k} \neq t_i \\
        0, & \text{otherwise}
    \end{cases}
\end{equation}

\noindent This is known as the ``B-Spline'' basis after it's inventor de Boor \citep{deBoor78}. The power of basis splines
comes from the fact that one only has to do the somewhat-expensive interpolation once for each set of points at which the spline is evaluated. 
This provides a considerable computational speedup as each evaluation of the spline function becomes a simpler operation: a dot product of a 
matrix and a vector. This straightforward operation is also ideal for optimizations from the use of GPU accelerators, 
enabling our Markov chain Monte Carlo (MCMC) based analyses, often with hundreds of parameters, to converge in an hour or less. 
Basis splines can easily be generalized to their two-dimensional analog, producing tensor product basis splines that, 
with this computational advantage, allow for high fidelity modeling of two-dimensional spline functions.

\begin{figure}[ht!]
    \begin{centering}
        \includegraphics[width=\linewidth]{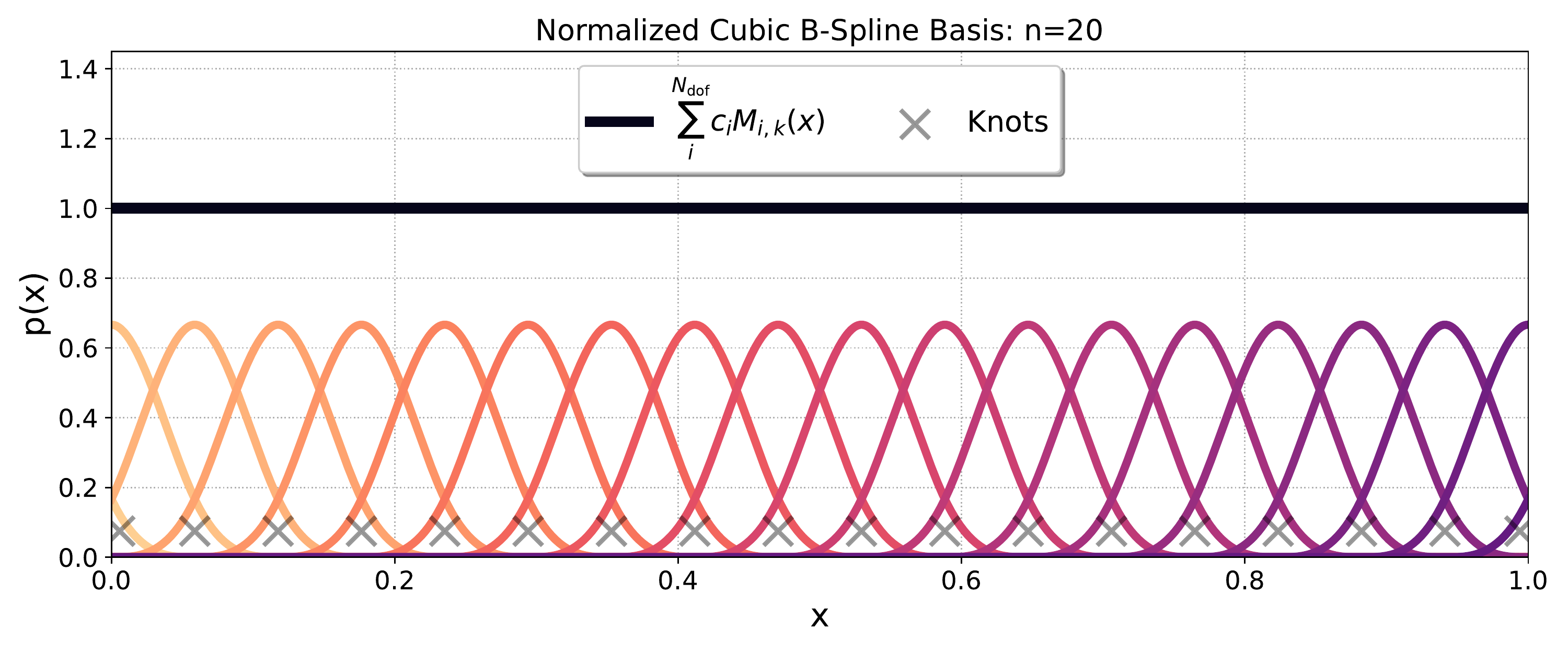}
        \caption{Plot showing a ``proper'' (see appendix \ref{sec:psplines}) normalized B-Spline basis of order 3 (cubic) with 20 degrees of freedom and equal weights for each component. 
        In black, we show the resulting spline function given equal weights and denote the location of the knots with gray x's. \LinkExplainer}
        \label{fig:spline_basis}
    \end{centering}
    \script{spline_basis_plot.py}
\end{figure}

Another important feature of basis splines is that under appropriate prior conditions, one can alleviate sensitivities to arbitrarily 
chosen prior specifications that splines commonly struggle with. Previous studies using splines had to perform multiple analyses, varying the 
number of spline knots, then either marginalized over the models or used model comparisons to motivate the best choice \citep{Edelman_2022ApJ}. 
We can avoid this step with the use of penalized splines (or P-Splines) \citep{eilers2021practical,BayesianPSplines,Jullion2007RobustSO}, 
where one adds a smoothing prior comprised of Gaussian distributions on the differences between neighboring basis spline coefficients. 
This allows for knots to be densely populated across the domain without the worry of extra variance in the inferred spline functions. 
When also fitting the scale of the smoothing prior (i.e. the width of the Gaussian distributions on the differences), the data will inform the model 
of the preferred the scale of smoothing required. We discuss the details of our smoothing prior implementation in more detail in the next section, 
Appendix \ref{sec:psplines}, following with our specific prior and basis choices for each model in Appendix \ref{sec:modelpriors}.

\section{Penalized Splines and Smoothing Priors}\label{sec:psplines}

Spline functions have been shown to be sensitive to the chosen number of knots, and their locations or spacing \citep{deBoor78}. 
Adding more knots increases the a priori variance in the spline function, while the space between knots can limit the 
resolution of features in the data the spline is capable of resolving. To ensure your spline based model is flexible enough 
one would want to add as many knots as densely as possible, but this comes with unwanted side effect of larger variance imposed by your model. 
This can be fixed with the use of penalized splines (P-Spline) in which one applies a prior or regularization term 
to the likelihood based on the difference of adjacent knot coefficients \citep{eilers2021practical}. The linear combination of spline basis components 
or the resulting spline function is flat when the basis coefficients are equal (see Figure \ref{fig:spline_basis}). By penalizing the likelihood as 
the differences between adjacent knot coefficients get larger, one gets a smoothing effect on the spline function \citep{eilers2021practical}. 
With hierarchical Bayesian inference as our statistical framework, we formulate the penalized likelihood of \citet{eilers2021practical}'s P-Splines with 
their Bayesian analog \citep{BayesianPSplines}. The Bayesian P-Spline prior places Gaussian distributions over the $r$-th order differences of the 
coefficients \citep{BayesianPSplines,Jullion2007RobustSO}. This is also sometimes referred to as a Gaussian random walk prior, and is similar in spirit to a Gaussian process prior used to regularize or smooth histogram bin heights as done in other non-parametric population studies \citep{Mandel_2016,o3b_astro_dist}. 
For a spline basis with $n$ degree's of freedom, and a difference penalty of order of $r$ (see \citet{eilers2021practical}), 
the smoothing prior on our basis spline coefficients, $\bm{c}$ is defined as:

\begin{eqnarray}
\bm{c} \sim \mathcal{N}(0, \sigma) \\
p(\bm{c} | \tau_\lambda) \propto \exp \big[ -\frac{1}{2} \tau_\lambda \bm{c}^{\mathrm{T}} \bm{D}_{r}^{\mathrm{T}} \bm{D}_r \bm{c}  \big] 
\end{eqnarray}

\noindent Above $\bm{D}_r$ is the order-$r$ difference matrix, of shape $(n-r \times n)$, and $\mathcal{N}(0,\sigma)$ a Gaussian distribution with zero mean 
and standard deviation, $\sigma$. This smoothing prior removes the strong dependence on number and location of knots that arises with using splines. 
The $\tau_\lambda$ controls the ``strength'' of the smoothing, or the inverse variance of the Gaussian priors on knot differences. We 
place uniform priors on $\tau_\lambda$ marginalize over this smoothing scale hyperparameter to let the data inform the optimal scale needed.
When there are a very large number of knots, such that your domain is densely populated with basis coefficients, this allows the freedom for the model to find the smoothing 
scale that the data prefers. 

This prior is imparting a natural attraction of the coefficients closer to each other in order to smooth the spline function, so one 
must ensure that the spline function is in fact flat given all equal coefficients. There needs to be $n+k+1$ knots to construct an order-k 
basis with n degrees of freedom. Some studies place knots on top of each other at hard parameter boundaries \citep{deBoor78,monotone_regression_splines}, 
which may seem motivated, but this violates the above condition necessary for the P-Spline prior. We follow the distinction in \citet{eilers2021practical} 
that such a smoothing prior is only valid with ``proper'' spline bases. A proper basis is where all $n+k+1$ knots are evenly and equally spaced, 
see Figure \ref{fig:spline_basis}, as opposed to stacking them at the bounds.

\section{Hierarchical Bayesian Inference} \label{sec:hierarchical_inference}

We use hierarchical Bayesian inference to infer the population properties of compact binaries. We want to infer the number density of merging compact binaries  
in the universe and how this can change with their masses, spins, etc. Often times it is useful to formulate the question in terms of the 
merger rates which is the number of mergers per $Gpc^{3}$ co-moving volume per year. For a set of hyperparameters, $\Lambda$, $\lambda$, and overall 
merger rate, $\mathcal{R}$, we write the overall number density of BBH mergers in the universe as: 

\begin{equation} \label{number_density}
     \frac{dN(\theta, z | \mathcal{R}, \Lambda, \lambda)}{d\theta dz} = \frac{dV_c}{dz}\bigg(\frac{T_\mathrm{obs}}{1+z}\bigg) \frac{d\mathcal{R}(\theta, z | \mathcal{R}_0, \Lambda, \lambda)}{d\theta} = \mathcal{R} p(\theta | \Lambda) p(z | \lambda)
\end{equation}

\noindent
where up above, we denote the co-moving volume element as $dV_c$ \citep{hogg_cosmo}, and $T_\mathrm{obs}$ as the observing time period that produced the 
catalog with the related factor of $1+z$ converting this detector-frame time to source-frame. We assume a Lambda CDM cosmology using 
the cosmological parameters from \citet{Planck2015}. We model the merger rate evolving with redshift following a power law distribution: 
$p(z|\lambda) \propto \frac{dV_c}{dz}\frac{1}{1+z}(1+z)^\lambda$ \citep{Fishbach_2018redshift}. When integrating equation \ref{number_density} across all $\theta$
and out to some maximum redshift, $z_\mathrm{max}$, we get the total number of compact binaries in the universe out to that redshift. We follow previous notations, \
letting $\{d_i\}$ represent the set of data from $N_\mathrm{obs}$ compact binaries observed with gravitational waves. The merger rate is then described as an inhomogeneous 
Poisson process and after imposing the usual log-uniform prior on the merger rate, we marginalize over the merger rate, $\mathcal{R}$, and arrive at the posterior
distribution of our hyperparameters, $\Lambda$ \citep{Mandel_2019, Vitale_2021}.

\begin{equation}
    p\left(\Lambda, \lambda | \{d_i\}\right) \frac{p(\Lambda)p(\lambda)}{\xi(\Lambda,\lambda)^{N_\mathrm{obs}}} \prod_{i=1}^{N_\mathrm{obs}} \bigg[ \frac{1}{K_i} \sum_{j=1}^{K_i} \frac{p(\theta^{i,j}|\Lambda)p(z^{i,j}|\lambda)}{\pi(\theta, z^{i,j})} \bigg]
\end{equation}

\noindent
where above, we replaced the integrals over each event's likelihood with ensemble averages over $K_i$ posterior samples \citep{GWTC3DATA}. Above, $j$
indexes the $K_i$ posterior samples from each event and $\pi(\theta, z)$ is the default prior used by parameter estimations that 
produced the posterior samples for each event. In the analyses of GWTC-3, either the default prior used was uniform in detector frame masses, 
component spins and Euclidean volume or the posterior samples were re-weighted to such a prior before using them in our analysis. 
The corresponding prior evaluated in the parameters we hierarchically model, i.e. source frame primary mass, mass ratio, component spins and redshift is:

\begin{equation}
    \pi(m_1, q, a_1, a_2, \cos{\theta_1}, \cos{\theta_2}, z) \propto \frac{1}{4} m_1 (1+z)^2 D_L^2(z) \frac{dD_L}{dz}
\end{equation}

\noindent Above, $D_L$ is the luminosity distance. To carefully incorporate selection effects to our model we need to quantify the detection efficiency,
$\xi(\Lambda, \lambda)$, of the search pipelines that were used to create GWTC-3, at a given population distribution described by $\Lambda$ and $\lambda$.
 
\begin{equation}
     \xi(\Lambda, \lambda) = \int d\theta dz P_\mathrm{det}(\theta, z)p(\theta | \Lambda) p(z | \lambda)
\end{equation}
 
\noindent
To estimate this integral we use a software injection campaign where gravitational waveforms from a large population of simulated sources. 
These simulated waveforms are put into real detector data, and then this data is evaluated with the same search pipelines that were used to 
produce the catalog we are analyzing. With these search results in hand, we use importance sampling and evaluate the integral 
with the Monte Carlo sum estimate $\mu$, and its corresponding variance and effective number of samples:

\begin{equation} \label{xi}
     \xi(\Lambda, \lambda) \approx \mu(\Lambda, \lambda) \frac{1}{N_\mathrm{inj}} \sum_{i=1}^{N_\mathrm{found}} \frac{p(\theta^i | \Lambda) p(z^i | \lambda)}{p_\mathrm{inj}(\theta, z^i)}
\end{equation}

\begin{equation}
    \sigma^2(\Lambda, \lambda) \equiv \frac{\mu^2(\Lambda, \lambda)}{N_\mathrm{eff}} \simeq \frac{1}{N^2_\mathrm{inj}} \sum_{i=1}^{N_\mathrm{found}} \bigg[\frac{p(\theta | \Lambda) p(z | \lambda)}{p_\mathrm{inj}(\theta, z)}\bigg]^2 - \frac{\mu^2(\Lambda, \lambda)}{N_\mathrm{inj}}
\end{equation}

\noindent
where the sum is only over the $N_\mathrm{found}$ injections that were successfully detected out of $N_\mathrm{inj}$ total injections, 
and $p_\mathrm{inj}(\theta, z)$ is the reference distribution from which the injections were drawn. We use the LVK released injection sets that describe the 
detector sensitivities over the first, second and third observing runs \citep{O1O2O3InjectionSets}. Additionally, we follow the procedure 
outlined in \citet{Farr_2019} to marginalize the uncertainty in our estimate of $\xi(\Lambda, \lambda)$, in which we verify that $N_\mathrm{eff}$ is 
sufficiently high after re-weighting the injections to a given population (i.e. $N_\mathrm{eff} > 4N_\mathrm{obs}$). 
The total hyper-posterior marginalized over the merger rate and the uncertainty in the Monte Carlo integral calculating $\xi(\Lambda, \lambda)$ \citep{Farr_2019}, as:

\begin{equation}\label{importance-posterior}
    \log p\left(\Lambda, \lambda | \{d_i\}\right) \propto \sum_{i=1}^{N_\mathrm{obs}} \log \bigg[ \frac{1}{K_i} \sum_{j=1}^{K_i} \frac{p(\theta^{i,j}|\Lambda)p(z^{i,j}|\lambda)}{\pi(\theta^{i,j}, z^{i,j})} \bigg] -  \\
    N_\mathrm{obs} \log \mu(\Lambda, \lambda) + \frac{3N_\mathrm{obs} + N_\mathrm{obs}^2}{2N_\mathrm{eff}} + \mathcal{O}(N_\mathrm{eff}^{-2}).
\end{equation}

We explicitly enumerate each of the models used in this work for $p(\theta|\Lambda)$, along with 
their respective hyperparameters and prior distributions in the next section. To calculate draw 
samples of the hyperparameters from the hierarchical posterior distribution shown in equation \ref{importance-posterior}, we use the 
NUTS Hamiltonian Monte Carlo sampler in \textsc{NumPyro} and \textsc{Jax} to calculate likelihoods \citep{jax,pyro,numpyro}.

\begin{table*}[b!]
    \centering
    \begin{tabular}{|l|l|l|l|}
    \hline
    \textbf{Model} & \textbf{Parameter} & \textbf{Description} & \textbf{Prior} \\ \hline \hline
    \multicolumn{4}{|c|}{\textbf{Primary Mass Model Parameters}} \\ \hline
    \textsc{B-Spline Primary} & $\bm{c}$ & Basis coefficients & $\sim \mathrm{Smooth}(\tau_\lambda, \sigma, r, n)$ \\ \cline{2-4} 
     & $\tau_\lambda$ & Smoothing Prior Scale & $\sim \mathrm{U}(2,1000)$ \\ \cline{2-4}
     & $r$ & order of the difference matrix for the smoothing prior & 2 \\ \cline{2-4} 
     & $\sigma$ & width of Gaussian priors on coefficients in smoothing prior & 6 \\ \cline{2-4} 
     & $n$ & number of knots in the basis spline & 64 \\ \hline \hline 
    \multicolumn{4}{|c|}{\textbf{Mass Ratio Model Parameters}} \\ \hline
    \textsc{B-Spline Ratio} & $\bm{c}$ & Basis coefficients & $\sim \mathrm{Smooth}(\tau_\lambda, \sigma, r, n)$ \\ \cline{2-4} 
     & $\tau_\lambda$ & Smoothing Prior Scale & $\sim \mathrm{U}(1,100)$ \\ \cline{2-4}
     & $r$ & order of the difference matrix for the smoothing prior & 2 \\ \cline{2-4} 
     & $\sigma$ & width of Gaussian priors on coefficients in smoothing prior & 4 \\ \cline{2-4} 
     & $n$ & number of knots in the basis spline & 18 \\ \hline \hline
    \multicolumn{4}{|c|}{\textbf{Redshift Evolution Model Parameters}} \\ \hline
    \textsc{PowerLaw+B-Spline} & $\lambda$ & slope of redshift evolution power law $(1+z)^\lambda$ &  $\sim \mathcal{N}(0,3)$ \\ \cline{2-4}
    & $\bm{c}$ & Basis coefficients & $\sim \mathrm{Smooth}(\tau_\lambda, \sigma, r, n)$ \\ \cline{2-4} 
     & $\tau_\lambda$ & Smoothing Prior Scale & $\sim \mathrm{U}(1,10)$ \\ \cline{2-4}
     & $r$ & order of the difference matrix for the smoothing prior & 2 \\ \cline{2-4} 
     & $\sigma$ & width of Gaussian priors on coefficients in smoothing prior & 1 \\ \cline{2-4} 
     & $n$ & number of knots in the basis spline & 18 \\ \hline \hline 
    \multicolumn{4}{|c|}{\textbf{Spin Distribution Model Parameters}} \\ \hline
    \textsc{B-Spline Magnitude} & $\bm{c}$ &  Basis coefficients & $\sim \mathrm{Smooth}(\tau_\lambda, \sigma, r, n)$  \\ \cline{2-4} 
    & $\tau_\lambda$ & Smoothing Prior Scale & $\sim \mathrm{U}(1,10)$ \\ \cline{2-4}
    & $r$ & order of the difference matrix for the smoothing prior & 2 \\ \cline{2-4} 
    & $\sigma$ & width of Gaussian priors on coefficients in smoothing prior & 1 \\ \cline{2-4} 
    & $n$ & number of knots in the basis spline & 18 \\ \hline \hline 
    \textsc{B-Spline Tilt} & $\bm{c}$ &  Basis coefficients & $\sim \mathrm{Smooth}(\tau_\lambda, \sigma, r, n)$  \\ \cline{2-4} 
    & $\tau_\lambda$ & Smoothing Prior Scale & $\sim \mathrm{U}(1,10)$ \\ \cline{2-4}
    & $r$ & order of the difference matrix for the smoothing prior & 2 \\ \cline{2-4} 
    & $\sigma$ & width of Gaussian priors on coefficients in smoothing prior & 1 \\ \cline{2-4} 
    & $n$ & number of knots in the basis spline & 18 \\ \hline \hline
    \end{tabular}
    \caption{All hyperparameter prior choices for each of the newly introduced basis spline models from this manuscript. See appendix 
    \ref{sec:basis_splines} and \ref{sec:psplines} for more detailed description of basis spline or smoothing prior parameters.}
    \label{tab:model_priors}
\end{table*} 

\section{Model and Prior Specification} \label{sec:modelpriors}

For each of the distributions with basis spline distributions, we have 2 fixed hyperparameters to specify. 
The number of degrees of freedom, $n$, and the difference 
penalty order for the smoothing prior, $r$. Additionally, one must choose a prior distribution on the smoothing prior scale hyperparameter, 
$\tau_\lambda$, which we take to be Uniform. For the primary mass distribution we model the log probability with a B-Spline interpolated 
in $\log(m_1)$ space. We follow a similar scheme for the models in mass ratio and spin, except we model the log probability with 
B-Splines that are interpolated in $q$, $a_i$ or $\cos{\theta_i}$ space. We adopt a minimum black hole mass of $5\msun$, 
and maximum of $100\msun$ with the equally spaced in this range. The knots for the mass ratio B-Spline are equally spaced 
from $\frac{m_\mathrm{min}}{m_\mathrm{max}}=0.05$ to $1$. There is motivation for the evolution of the merger rate with redshift 
to follow a power law form since it should be related to the star formation rate \citep{Madau_2014}, 
motivating our adoption of a semi-parametric approach where we use B-Splines to model modulations to the 
simpler underlying \textsc{PowerlawRedshift} model \citep{Fishbach_2018redshift,Edelman_2022ApJ}. 
We model modulations to the underlying probability density with the multiplicative factor, $e^{B(\log z)}$, 
where $B(\log z)$ is the B-Spline interpolated from knots spaced linearly in $\log z$ space. 
We enumerate each of our specific model hyperparameter 
and prior choices in table \ref{tab:model_priors}.

\section{Posterior Predictive Checks} \label{sec:ppcs}

\begin{figure}
    \includegraphics[width=\linewidth]{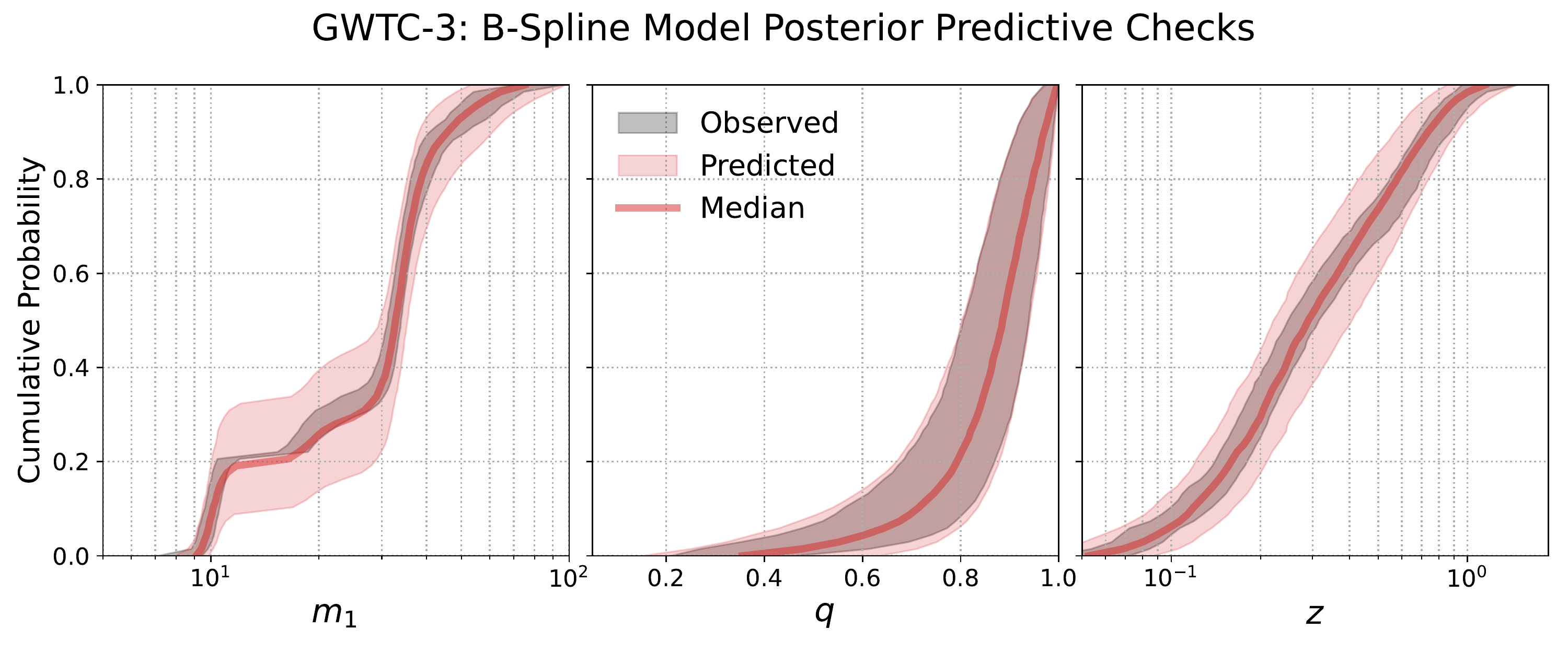}
    \caption{Posterior predictive checks showing the CDFs of the observed (black) and predicted (red) distributions of GWTC-3 sized catalogs for each posterior sample of the IID spin B-Spline model. The shaded regions show 90\% credible intervals and the solid red line is the median of the predicted distribution. \LinkExplainer}
    \label{fig:ppc}
    \script{plot_ppcs.py}
\end{figure}

\begin{figure}
    \includegraphics[width=\linewidth]{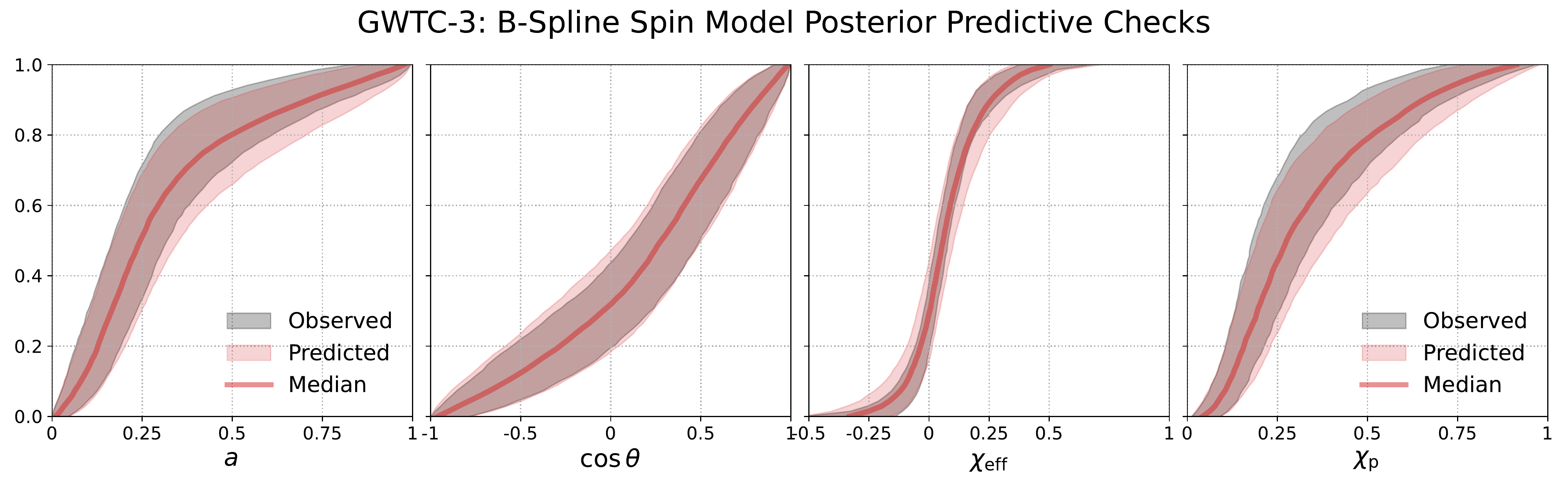}
    \caption{Posterior predictive checks showing the CDFs of the observed (black) and predicted (red) distributions of GWTC-3 sized catalogs for each posterior sample of the IID spin B-Spline model. The shaded regions show 90\% credible intervals and the solid red line is the median of the predicted distribution. \LinkExplainer}
    \label{fig:spin_ppc}
    \script{plot_spin_ppcs.py}
\end{figure}

We follow the posterior predictive checking procedure done in recent population studies to validate our models inferences \citep{o3a_pop,Edelman_2022ApJ}. 
For each posterior sample describing our model's inferred population we reweigh the observed event samples and the found injections to that population and draw a set 69 (size of GWTC-3 BBH catalog) samples to construct the observed and predicted distributions we show in figure \ref{fig:ppc} and figure \ref{fig:spin_ppc}. When the observed region stays encompassed within the predicted region the model is performing well, which we see across each of the fit parameters. 

\section{Reproducibility}
\label{sec:reproducibility}

In the spirit of open source and reproducible science, this study was done using the reproducibility software \href{https://github.com/showyourwork/showyourwork}{\showyourwork} \citep{Luger2021}, which leverages continuous integration to programmatically download the data from \href{https://zenodo.org/}{zenodo.org}, create the figures, and compile the manuscript. Each figure caption contains two links that point towards the dataset (stored on zenodo) used in the corresponding figure, and to the script used to make the figure (at the commit corresponding to the current build of the manuscript). The git repository associated to this study is publicly available at \url{https://github.com/bruce-edelman/CoveringYourBasis}, which allows anyone to re-build the entire manuscript. The datasets and all analysis or figure generating scripts are all stored on \href{https://zenodo.org/}{zenodo.org} at \url{https://zenodo.org/record/7566301} \citep{edelman_bruce_2022_7422514}.